\documentclass[12pt,preprint]{aastex}
\usepackage{psfig}
\usepackage{graphics}

\shorttitle{Dust and Water in IRAS 16293$-$2422}
\shortauthors{Stark et al.}

\newcommand{\Msun}{\,{\rm M$_{\odot}$}}
\newcommand{\Lsun}{\,{\rm L$_{\odot}$}}
\newcommand{\asec}{\,{$''$}}
\newcommand{\amin}{\,{$'$}}
\newcommand{\degree}{{$^{\circ}$}}
\newcommand{\kms}{\,{{km~s$^{-1}$}}}
\newcommand{\mic}{\mbox{\,${\mu}$m}}

\begin{document}

\title{Probing the Early Stages of Low-Mass Star Formation in
  LDN~1689N:\\ Dust and Water in IRAS 16293$-$2422A, B, and E}

\author{Ronald Stark}
\affil{Max-Planck-Institut f\"ur Radioastronomie, Auf dem H\"ugel 69, 
  D-53121 Bonn, Germany}
\affil{Sterrewacht Leiden, P.O. Box 9513, 2300 RA Leiden, The Netherlands}

\author{G\"oran Sandell}
\affil{NASA Ames Research Center, MS 144-2, Moffett Field, CA 94035}

\author {Sara C. Beck }
\affil{Department of Physics and Astronomy, Tel Aviv University, Ramat
  Aviv, Israel}
\affil{Harvard-Smithsonian Center for Astrophysics, 60 Garden Street,
  Cambridge, MA 02138}

\author{Michiel R. Hogerheijde} \affil{Steward Observatory, The
University of Arizona, 933 N. Cherry Ave, Tucson, AZ
85721-0065\footnote{Current Address: Sterrewacht Leiden, P.O. Box
9513, 2300 RA Leiden, The Netherlands}}

\author{Ewine F. van Dishoeck}
\affil{Sterrewacht Leiden, P.O. Box 9513, 2300 RA Leiden, The Netherlands}

\author{Peter van der Wal, Floris F.~S. van der Tak, Frank Sch\"afer}
\affil{Max-Planck-Institut f\"ur Radioastronomie, Auf dem H\"ugel 69, 
  D-53121 Bonn, Germany}

\author {Gary J. Melnick, Matt L.~N. Ashby}
\affil{Harvard-Smithsonian Center for Astrophysics, 60 Garden Street,
Cambridge, MA 02138}

\and

\author{Gert de Lange}
\affil{Space Research Organisation of the Netherlands (SRON), 
  P.O. Box 800, 9700 AV Groningen, The Netherlands}

%\newpage

\begin{abstract}
We present deep images of dust continuum emission at 450, 800, and
850~$\mu$m of the dark cloud LDN~1689N which harbors the low-mass
young stellar objects (YSOs) IRAS~16293$-$2422A and B (I16293A and
I16293B) and the cold prestellar object I16293E. Toward the positions
of I16293A and E we also obtained spectra of CO-isotopomers and deep
submillimeter observations of chemically related molecules with high
critical densities (HCO$^+$, H$^{13}$CO$^+$, DCO$^+$, H$_2$O, HDO,
H$_2$D$^+$). To I16293A we report the detection of the HDO
$1_{01}$--$0_{00}$ and H$_2$O $1_{10}$--$1_{01}$ ground-state
transitions as broad self-reversed emission profiles with narrow
absorption, and a tentative detection of H$_2$D$^+$ $1_{10}$--$1_{11}$.
To I16293E we detect weak emission of subthermally excited HDO
$1_{01}$--$0_{00}$. Based on this set of submillimeter continuum and
line data we model the envelopes around I16293A and E. The density and
velocity structure of I16293A is fit by an inside-out collapse model,
yielding a sound speed of $a$=0.7~km~s$^{-1}$, an age of
$t$=(0.6--2.5)$\times 10^4$~yr, and a mass of 6.1~M$_\sun$. The
density in the envelope of I16293E is fit by a radial power law with
index $-1.0\pm 0.2$, a mass of 4.4~M$_\sun$, and a constant
temperature of 16~K. These respective models are used to study the
chemistry of the envelopes of these pre- and protostellar objects.

We made a large, fully sampled CO $J$=2--1 map of LDN~1689N which
clearly shows the two outflows from I16293A and B, and the interaction
of one of the flows with I16293E. An outflow from I16293E as reported
elsewhere is not confirmed. Instead, we find that the motions around
I16293E identified from small maps are part of a larger scale fossil
flow from I16293B. Modeling of the I16293A outflow shows that the
broad HDO, water ground-state, and CO $J$=6--5 and 7--6 emission lines
originate in this flow, while the HDO and H$_2$O line cores originate
in the envelope. The narrow absorption feature in the ground-state
water line is due to cold gas in the outer envelope. The derived
H$_2$O abundance is $3 \times 10^{-9}$ in the cold regions of the
envelope of I16293A ($T_{\rm kin}<14$~K), $2\times 10^{-7}$ is warmer
regions of the envelope ($>14$~K), and $10^{-8}$ in the outflow. The
HDO abundance is constant at a few $\times 10^{-10}$ throughout the
envelopes of I16293A and E. Because the derived H$_2$O and HDO
abundances in the two objects can be understood through shock
chemistry in the outflow and ion-molecule chemistry in the envelopes,
we argue that both objects are related in chemical evolution. The
[HDO]/[H$_2$O] abundance ratio in the warm inner envelope of I16293A of
a few times 10$^{-4}$ is comparable to that measured in comets. This
supports the idea that the [HDO]/[H$_2$O] ratio is determined in the
cold prestellar core phase and conserved throughout the formation
process of low-mass stars and planets.
 \end{abstract}


\section{Introduction\label{s:intro}}

IRAS~16293$-$2422 is one of the best studied low-mass young stellar
objects (YSOs) \citep{bla94,evd95,cec00}. It is deeply embedded in the
LDN~1689N cloud in Ophiuchus (distance 160~pc) and classified as an
extreme Class I/Class 0 YSO \citep{Lad91}. It is a low-luminosity
(27~$L_\sun$, \citealt{Walk86}) binary system (hereafter, I16293A and
I16293B) with a separation of 840~AU ($5''$) and individual stellar
masses of 0.5~M$_\sun$ \citep{Mun92}. Two powerful outflows emanate
from the system.  At least three physically and chemically different
regions can be recognised in a $20''$ (3000~AU) beam \citep{evd95}:
(i) a cold, relatively low-density outer envelope ($T_{\rm kin} \simeq
$10--20~K, $n$(H$_2)\simeq 10^4 $--$10^5$ cm$^{-3}$) which marks the
transition into the extended parental cloud LDN~1689N; (ii) a warmer
circumbinary envelope of size $\sim 2000$~AU ($T_{\rm kin}\simeq
40$~K, $n$(H$_2)\simeq 10^6$--$10^7$~cm$^{-3}$); and (iii) a warm,
dense core of about 500--1500~AU ($T_{\rm kin} \ge 80$ K,
$n$(H$_2)\simeq 10^7$ cm$^{-3}$) which traces the interaction of the
outflow and the stellar radiation with the inner part of the
circumbinary envelope. This picture emerged from excitation analysis
of species tracing these distinct regions and conditions, assuming
uniform temperatures and densities for each. Continuous descriptions
of the density and temperature as function of radius in the collapsing
envelope were developed by \citet{cec00} and \citet{Scho02} on scales
from 30 to 5000~AU. In this Paper, we present new observations and
modeling, that allows us to independently derive the density and
temperature structure in the extended envelope around the (pre-)
protostellar cores through continuum photometry and imaging of the
dust, and heterodyne spectroscopy of the molecular gas. The resulting
density and temperature distribution ranges from the warm inner
envelope region (100~AU) to the coldest (12~K) outer envelope regions
(7300~AU) and serves as a basis for excitation and abundance analysis
of the line data.

Another component of the IRAS~16293$-$2422 system is located 80$''$
east and 60$''$ south of I16293A. This small clump, I16293E, has
strong DCO$^+$, NH$_3$, and NH$_2$D emission \citep{Woo87,Mun90,Sha01}
connected to the I16293A core, and significant submillimeter continuum
emission from dust \citep{San94}. Single-dish observations of NH$_3$
in a $40\arcsec$ beam indicate a temperature $T_{\rm kin} = 12$~K
\citep{Miz90}. Millimeter-interferometric data, sampling smaller
scales and less sensitive to the warmer extended envelope, yield
$T_{\rm kin} \simeq 8$~K and $n({\rm H}_2)\simeq 1\times 10^5$
cm$^{-3}$ if homogeneous conditions are assumed \citep{Sha01}.
Because no far-infrared (FIR) or submillimeter point source has been
found embedded in I16293E, it is classified as a pre-(proto)stellar
core. In this Paper, we present submillimeter images that constrain
the density structure of I16293E.

It is difficult to study chemical evolution in the early protostellar
stages, because during collpase the temperature stays low at $\sim
10$~K while the density increases to $n($H$_2)\ge 10^8$ cm$^{-3}$.
Under these conditions the line emission of many molecules is
dominated by a warmer outer envelope around the prestellar core. Also,
many molecules in the cold core will condense on dust
grains. Eventually, only H$_2$, H$_3^+$, and their isotopomers HD and
H$_2$D$^+$ will remain in the gas phase. This leads to a significant
enhancement of H$_2$D$^+$ through the deuterium exchange reaction
\begin{equation}               % EQUATION 1: H3+/H2D+ balance
{\rm H}_3^+ +{\rm  HD} 
   \rightleftharpoons {\rm H}_2{\rm D}^+ + {\rm H}_2 + \Delta E,
\label{e1}
\end{equation} 
since the backward reaction becomes negligible at low temperatures
\citep{Smi82,Her82}. H$_2$D$^+$ is thought to play a pivotal role in
the deuteration of molecules \citep{Wat76,Mil89} and the detection of
H$_2$D$^+$ in the Class~0 YSO NGC~1333 IRAS~4A \citep{Sta99} is an
important confirmation of the cold gas-phase chemical networks. The
H$_2$D$^+$ enhancement is reflected in the high abundance ratios that
are generally observed of, e.g., [DCO$^+$]/[HCO$^+$],
[NH$_2$D]/[NH$_3$], [DCN]/[HCN], and [N$_2$D$^+$]/[N$_2$H$^+$] in cold
prestellar cores and Class~0 YSOs (e.g., \citealt{Gue82, Olb85,
Woo87a, But95, Wil98, Ber02}). Models predict that in the collapse
stage the abundances of many deuterated radicals and molecules
increase sharply after their non-deuterated versions start to get
heavily depleted on dust grains, then reach a peak and start to
decrease as the deuterated molecules too condense on the dust (e.g.,
\citealt{Bro89, Rod96}). Reactions with atomic deuterium on the grain
surface may further enhance the abundance of deuterated species in the
ice \citep{Tie89,Bro89}. After the formation of a YSO, the condensed
molecules may evaporate from grains, resulting in a temporary increase
in the gas-phase deuteration fraction. The deuteration may decline
after $\sim 10^4$ years \citep{Rod96} when higher temperature
chemistry becomes effective. In addition, in such hot ($T_{\rm kin}
\simeq 100$--200 K) regions the reversal of reaction (\ref{e1})
becomes dominant and very little new fractionation is expected to
occur. We report in this Paper observations of several key deuterated
species in I16293A and I16293E.

Water is an important species in the early stages of star
formation. H$_2$O is the most important hydride molecule in oxygen
chemistry and models predict that H$_2$O plays a dominant role in the
energy balance and chemical evolution during star formation. Water
will have the largest abundance and excitation contrast between the
protostellar source and the surrounding cloud (e.g., \citealt{Cec96,
Dot97}). Measurements of the rotational and ro-vibrational transitions
of H$_2$O in star-forming regions have recently become available from
the Infared Space Observatory (ISO; e.g., \citealt{Hel96a, cec00}) and
the Submillimeter Wave Astronomical Satellite (SWAS; e.g.,
\citealt{Ash00, Sne00, Neu00}). In particular the heterodyne
velocity-resolved measurements of the ground-state transition of
ortho-H$_2$O with SWAS allow a direct determination of the water
abundance throughout the envelope. The HDO $1_{01}$--$0_{00}$
ground-state transition at 465~GHz ($E_u= 23.2$ K, $n_{\rm crit} \ge
10^8$ cm$^{-3}$ for $T_{\rm kin} \le 50$ K, \citealt{Gre89}) is an
excellent tracer of high density gas at low temperatures where the
abundance of HDO relative to water is enhanced.  Ground-based
observations of the HDO ground-state transition are still very
sparse. To date, this transition has only been observed in the
high-mass star-forming regions Orion-KL \citep{Sch91}, W3(OH), and
W3(H$_2$O) \citep{Hel96b}. The latter study indicates that the
[HDO]/[H$_2$O] abundance ratio is comparable to that found in hot
cores and is not a sensitive indicator of the evolutionary stage in
high-mass star formation. This could mean either that the W3 cloud
always stayed warm or that the [HDO]/[H$_2$O] ratio retrusn to thermal
equilibrium faster than ratios like [DCN]/[HCN] \citep{Hel96b}. For a
low-mass Class~0 YSO like I16293A the situation may be different,
because the time scale to reach steady-state may be much longer. In
this Paper, we present observations of HDO and H$_2$O of I16293A and
E.

%% The physical parameters in cold (pre-) protostellar cores cannot
%% easily be derived from molecular line observations alone because of
%% the earlier mentioned excitation and depletion processes. Thus, even
%% the total hydrogen column density cannot be estimated accurately, and
%% the commonly adopted CO-isotopomer tracer method may underestimate
%% $N$(H$_2$) by more than a factor 10--20 (e.g., Blake et al. 1995; this
%% paper). This may also be true for IRAS 16293$-$2422 where most
%% previous abundance determinations with respect to H$_2$ are based on
%% single dish C$^{17}$O observations (e.g., Blake et al. 1994, van
%% Dishoeck et al. 1995, Loinard et al. 2001). Dust continuum photometry
%% and mapping at submillimeter wavelengths provide a more reliable way
%% to measure the density and mass of the dust and gas. Such measurements
%% also yield the emissivity and temperature of the cold dust and provide
%% information on the dust-to-gas coupling. Sch\"oier et al. (2002) have
%% recently reanalyzed the data of Blake et al. (1994) and van Dishoeck
%% et al. (1995) to determine the abundances of a variety of other
%% species using this method.

In this paper we present a detailed study of the physical and chemical
structure of the low-mass star-forming cloud LDN~1689N with particular
emphasis on the H$_2$O/HDO chemistry and deuterium fractionation.  We
present extensive submillimeter continuum photometry and maps
revealing the warm and cold dust around I16293A, B and I16293E (\S
\ref{Obs_cont}). We report the detection of the ground-state lines of
(deuterated) water and a tentative detection of H$_2$D$^+$ toward
I16293A as well as weak HDO emission in the prestellar core I16293E
(\S \ref{Obs_line}). These data are combined with pointed observations
of CO and HCO$^+$ isotopomers to determine the temperature and density
structure throughout the envelope (\S \ref{Results}). We address the
two outflows and their origin in \S \ref{CO_outflows}. Our deep
observations of the HDO and H$_2$O ground-state transitions are used
to study the [HDO]/[H$_2$O] ratio throughout the envelope (\S
\ref{hdo_h2o}).  A simple chemistry network is used to model the
deuterium chemistry and in particular the water fractionation (\S
\ref{Simple_chem_model}).  Finally, we discuss the evolutionary
difference between I16293A, I16293B, and I16293E (\S \ref{Nature_of}),
and summarize the conclusions in \S \ref{conclusion}.

\section{Observations and Data Reduction\label{observations}}

\subsection{Submillimeter Continuum\label{Obs_cont}}

All continuum observations were made with the James Clerk Maxwell
Telescope (JCMT), using the single channel bolometer UKT14 and the
versatile Submillimetre Common-User Bolometer Array (SCUBA) under good
weather conditions. The UKT14 measurements were done during the
winters of 1991--1993. This instrument was for a long time the
common-user bolometer system and is described in \citet{Dun90}. It was
replaced in 1997 by SCUBA which was designed both for photometry and
fast mapping \citep{Hol98}.

The photometry of I16293E was done with UKT14 during Spring 1993. The
chop throw was 90\asec\ in Declination and the calibrators were Uranus
and the nearby secondary calibration standard I16293A
\citep{San94}. The results are given in Table \ref{t1}. UKT14 maps
were obtained at 1.1 mm, 800\mic, 750\mic, 450\mic, and 350\mic\
during several observing runs in 1991 and 1992. Most of these maps
have poor sensivitity and were mainly used to derive the submillimeter
position of I16293A. However, even with the somewhat limited
signal-to-noise (S/N) these provide a better estimate of the
background subtracted flux of I$16293$A than single position
photometry.

\begin{figure} % Figure 1
  \epsscale{0.44}
\plotone{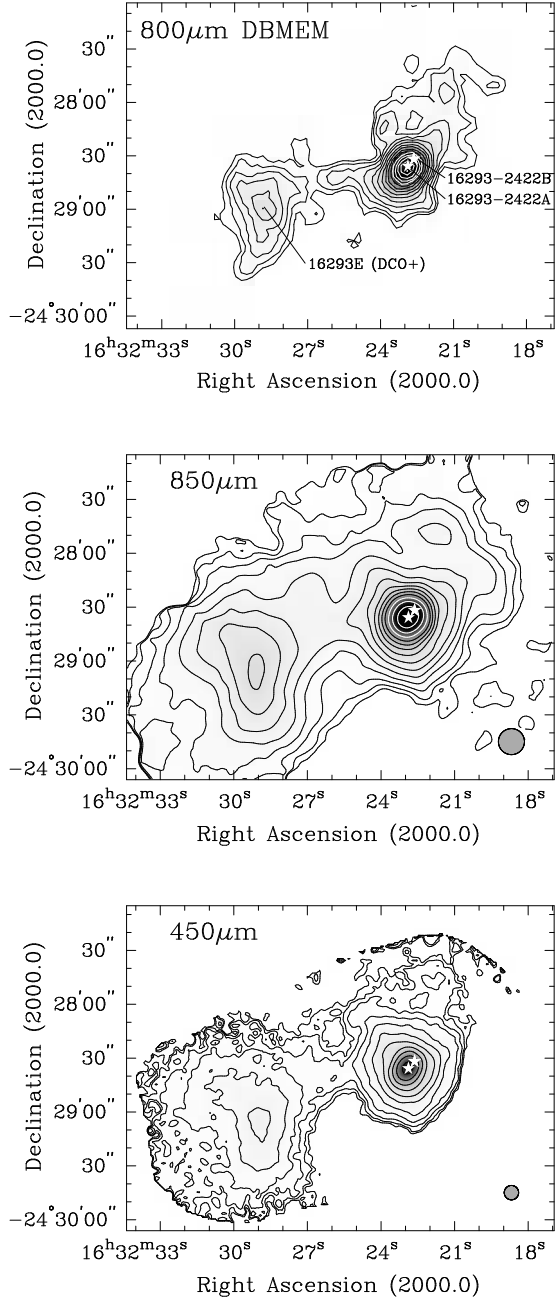}
  \caption{Continuum maps of IRAS 16293$-$2422A,\,B and I16293E, all
    shown in greyscale enhanced with logarithmic contours.  {\sl Top}:
    800\mic\ UKT14 map. The position of the core I16293E and the two
    protostars I16293A and B are indicated. The lowest contour is at
    0.1 Jy~pixel$^{-1}$ (3\asec\ pixels) and peak flux density is 1.8
    Jy~pixel$^{-1}$.  {\sl Middle}: 850\mic\ SCUBA map.  The lowest
    countour is at 0.1 Jy~beam$^{-1}$, peak flux is 16.0
    Jy~beam$^{-1}$.  {\sl Bottom}: 450\mic\ SCUBA map. The lowest
    contour is at 1 Jy~beam$^{-1}$, peak at 76.1 Jy~beam$^{-1}$. The
    excess noise seen in the outskirts of the map is not real but due
    to edge effects.  The effect from blanked negative bolometers
    north of I16293A,\,B is evident.  The HPBW is indicated for the
    850 and 450 \mic\ maps.\label{f1}}
\end{figure}

At 800\mic\ we attempted to go deep and completely map the dust cloud
surrounding the bright submillimeter source. Because the dust emission
is very extended (see the 850 $\mu$m SCUBA map in Fig. \ref{f1}) we
combined six maps of the region, centered on I16293A or I16293E. Each
map was bound by pointing observations of the nearby blazars
1514$-$241 and 1730$-$130. Typical pointing drifts were less than
2\asec. These errors were removed in the data reduction process
assuming that the pointing drifts linearly in Azimuth and Elevation as
a function of time. All maps were made on-the-fly, scanned in Azimuth
with a cell size of 4\asec\ and a chop throw of 40\asec\ in the scan
direction. With a single pixel instrument it is difficult to map such
an extended region. The final map was therefore reduced with the Dual
Beam Maximum Entropy (DBMEM) algorithm written by John Richer
\citep{Ric92} for JCMT. The DBMEM map was baseline subtracted and
recalibrated by using the integrated intensity in a $60'' \times 60''$
area around I16293A from maps reduced with NOD2
\citep{Has74,Em79}. The 800\mic-map is shown in Fig. \ref{f1} (top
panel). From comparison with SCUBA maps (see below), it is clear that
the 800\mic-map has failed to recover the faint, relatively uniform,
low-level emission surrounding both cloud cores. This does not affect
the morphology of the map, but does have a large effect on integrated
intensities of the surrounding LDN~1689N cloud. The low-level emission
is missing because the 800\mic-maps were taken with the same rather
short chop throw, and the individual maps have a noise level
comparable or higher to that of the low-level extended emission. It is
also well known that the NOD2-algorithm is not very sensitive to
relatively uniform extended emission on spatial scales of several
times the chop throw \citep{Em79}.

Additional maps were obtained with SCUBA during Spring 1997 as part of
commissioning jiggle- and scan-map observing modes. We have
complemented these maps with additional maps obtained from the JCMT
archive in the time period 1997--1998.  All maps were taken under
excellent sky conditions (atmospheric opacities $\tau_{225~{\rm GHz}}<
0.04$), and all the maps in the archive were taken with a 120\asec\
chop throw, typically in Azimuth. Most of the maps of I16293E were
also obtained with a 120\asec\ chop.

The pointing was checked before and after each map using the same
blazars as for our UKT14 observations. Some of the archive maps,
however, were taken without pointing observations. These were aligned
with the pointing corrected average before adding them into the final
data set.  Calibration and beam characterization is based on beam maps
of Uranus obtained with the same chop throw.  The Uranus maps give a
Half Power Beam Width of 14.5\asec\ and 7.8\asec\ for 850 $\mu$m and
450 $\mu$m, respectively.

Both the I16293A, B and I16293E cloud cores are rather extended and no
map is completely free of emission in the off positions. Any bolometer
that showed evidence for or was suspected of being contaminated by the
chop was therefore blanked in the data reduction.  Although this was
done as carefully as possible, it is clear that the outskirts of our
final mosaiced image (Fig. \ref{f1}) may still be affected,
especially since the dust emission extends beyond the area we mapped
to the north and northeast of both I16293A and I16293E. The cloud core
surrounding I16293E also extends further east than is covered by our
maps.

The basic data reduction was done using the SCUBA software reduction
package SURF \citep{Jen98} as explained in the SCUBA Map Reduction
Cookbook \citep{San01}. In total we used 16 data sets for the 850
$\mu$m map and 13 for the 450 $\mu$m map.  In the final coadd we
adjust the pointing in each map (shift and add) to ensure that the
final map is not broadened by small pointing errors. The rms noise
levels in the final maps are difficult to estimate, because there are
no emission-free regions in the maps, but we estimate them to be below
20 mJy~beam$^{-1}$ at 850 $\mu$m and below 80 mJy~beam$^{-1}$ at 450
$\mu$m.

The final maps were converted to FITS and exported to MIRIAD
\citep{Sau95} for further analysis. In order to correct for the error
beam contribution, especially at 450 $\mu$m, we deconvolved the maps
using clean and spherically symmetric model beams derived from beam
maps of Uranus observed in stable night time conditions. At 850 $\mu$m
we use HPBWs of 14.5\asec, 55\asec, and 150\asec, with amplitudes of
0.985, 0.014 and 0.001, respectively. At 450 $\mu$m the HPBWs are
7.8\asec, 34\asec, and 140\asec, with amplitudes of 0.964, 0.029, and
0.007, respectively.  The final maps were restored with a 14\asec\
beam at 850 and an 8\asec\ beam at 450 $\mu$m. The peak flux densities
in the restored maps are 16.0 Jy~beam$^{-1}$ and 76.1 Jy~beam$^{-1}$
toward I16293A at 850 and 450 $\mu$m, respectively, while the peak
fluxes of I16293E are 1.4 Jy~beam$^{-1}$ and 4.35 Jy~beam$^{-1}$ at
850 and 450 $\mu$m, respectively.

\subsection{Molecular line observations\label{Obs_line}}

\begin{figure} % Figure 2
  \epsscale{0.9}
\plotone{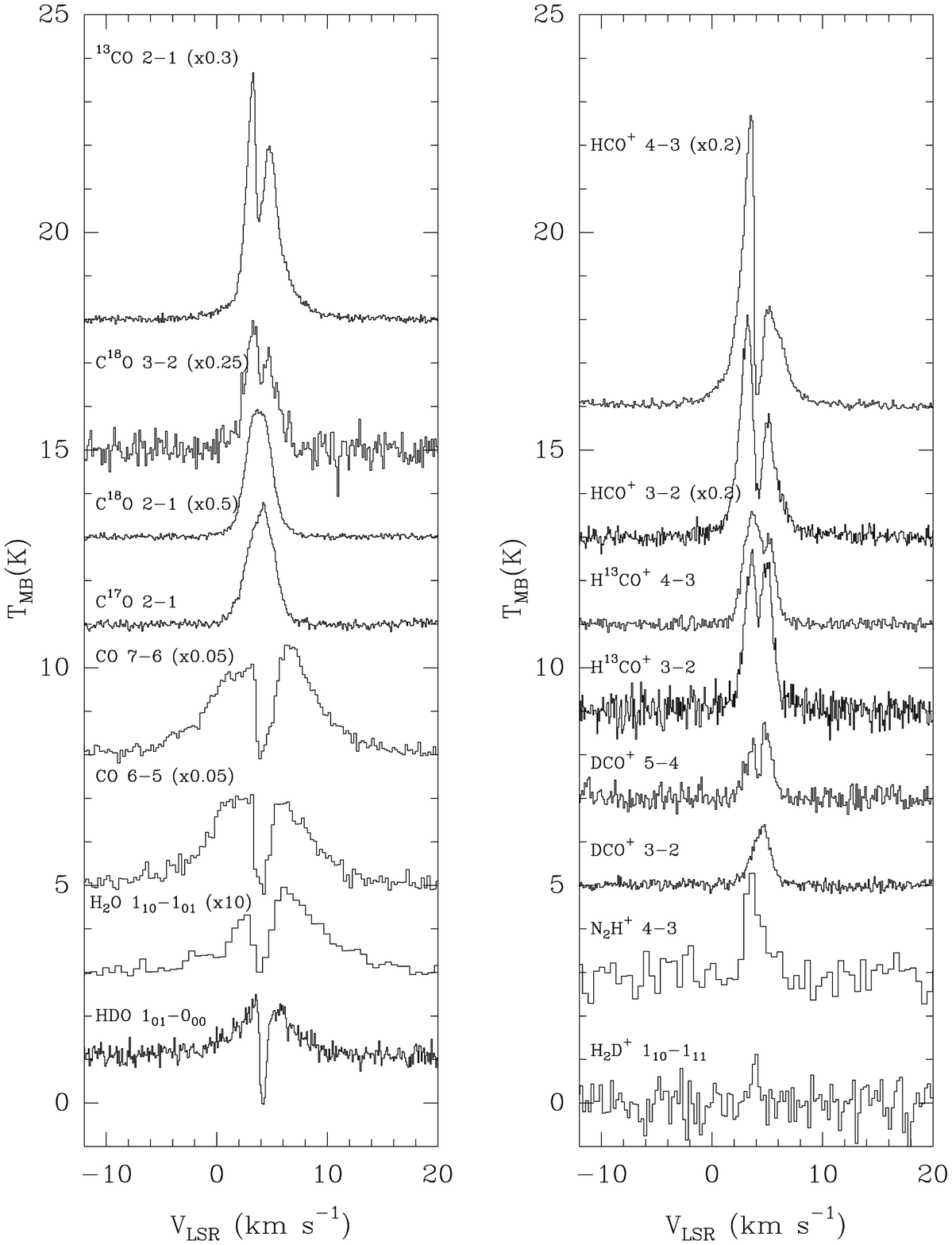}
  \caption{Observed spectra toward I16293A.\label{f2}}
\end{figure}

\begin{figure} % Figure 3
  \epsscale{0.4}
\plotone{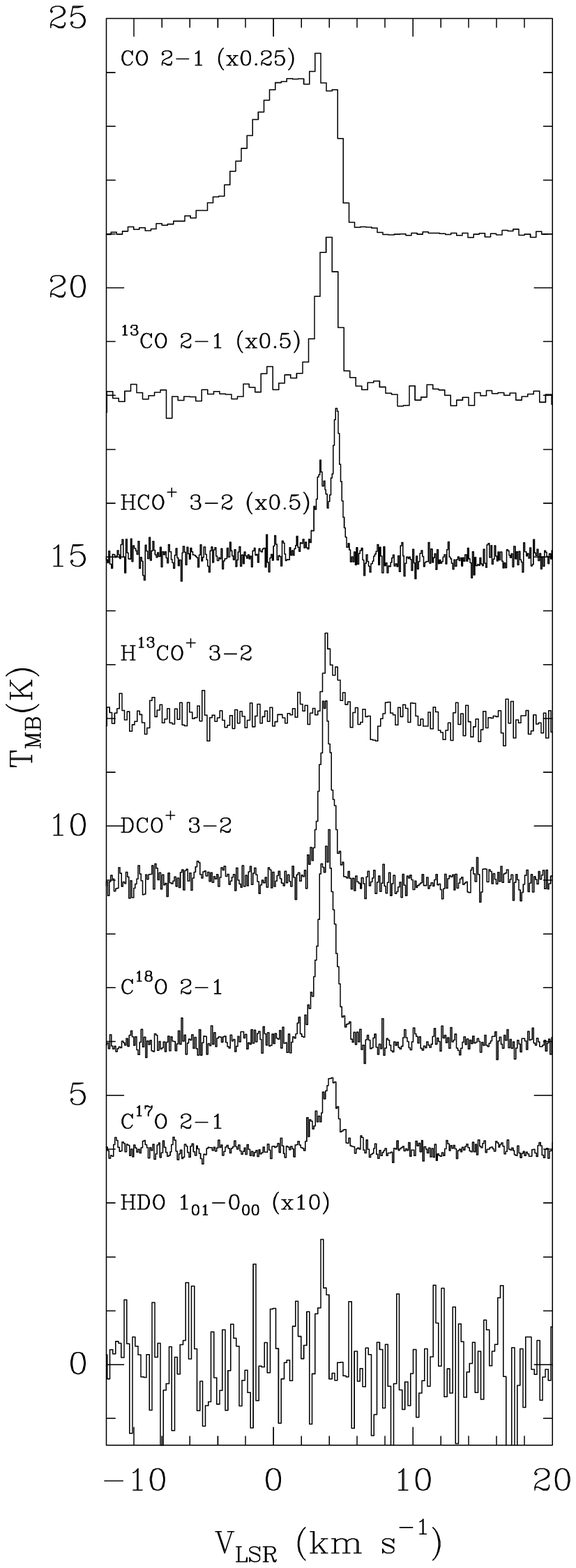}
  \figcaption{Observed spectra toward I16293E.\label{f3}}
\end{figure}

Spectra of CO $J$=2--1 (230.538000 GHz) and isotopomers, as well as
the $J$=3--2 transitions of DCO$^+$ (216.112605 GHz), HCO$^+$
(267.557619 GHz), and H$^{13}$CO$^+$ (260.255478 GHz) toward I16293A
and I16293E were retrieved and reduced from a search of all released
data in the JCMT archive at the Canadian Astronomy Data Center taken
with the SIS receiver RxA2 and its successor RxA3. Deep observations
of the HDO $3_{12}$--$2_{12}$ (225.896720 GHz) and $2_{11}$--$2_{12}$
(241.561550 GHz) lines were made with the receiver RxA3 in
2001~April. We also acquired a large, fully sampled map in the CO 2--1
line which is centered on I16293E and covers the outflows from I16293A
and B. The map was made on-the-fly with a bandwidth of 125~MHz, a
sampling of 5\arcsec\ in the scan direction, and a cross-scan step
size of 10\arcsec.  The final map was made from a series of sub-maps
scanned either in RA or Dec. The integration time per point was 5
seconds, but in poor weather conditions we often made multiple
coverages. All observations were done in position switch mode with the
reference position 600\asec\ or 800\asec\ west of the submillimeter
position of I16293E. At the same time we have also obtained spectra
toward I16293A, which is a pointing and spectral line standard. The
relative calibration of the spectra of the CO 2--1 map is accurate
within 5\%, because we overlapped the sub-maps. In 1999~April we
obtained additional deep spectra of CO and $^{13}$CO 2--1 (220.398677
GHz) at the peak positions in the outflow lobes in order to get more
accurate estimates of the $^{12}$CO optical depth in the high velocity
gas. Most of the spectra were obtained with a bandwidth of 125~MHz,
corresponding to a velocity resolution of 0.1\kms. The spectra are
calibrated in $T_{\rm MB}$ (I16293A) or $T_R^*$ (I16293E) and shown in
Figs. \ref{f2} and \ref{f3}. The results of gaussian fits to the line
profiles are summarised in Table \ref{t2}. For some of the lines with
symmetric profiles (e.g., HDO) the emission and the absorption
components were fitted separately. For most of the complex line
profiles we only list the integrated emission and the velocity of the
maximum self-absorption.  Note that the phase-lock instability of RxA2
and RxA3 may cause an instrumental broadening up to 0.5 km~s$^{-1}$.

The N$_2$H$^+$ $J$=4--3 (372.672509 GHz), H$_2$D$^+$
$1_{10}$--$1_{11}$ (372.421340 GHz), DCO$^+$ 5--4 (360.169881 GHz),
HCO$^+$ 4--3 (356.734288 GHz), and H$^{13}$CO$^+$ 4--3 (346.998540
GHz) rotational transitions were observed in 2000~April with the JCMT
using the dual channel SIS receiver RxB3. The receiver was operated in
single side band (SSB) mode where the image side band is rejected and
terminates in a cold load. The main beam efficiency at these
frequencies is about 60\%, while the beam width is 14--13\asec for
345--372 GHz. The lines were observed in beamswitching mode switching
over 180\asec. The N$_2$H$^+$ and H$_2$D$^+$ lines were observed
simultaneously with relatively good atmospheric transparency
($\tau_{225~{\rm GHz}}\simeq 0.05$), but with only 30 minutes of
useful integration time above an elevation of 40 degrees. RxB3 was
changed to tunerless mixers in 1999 and the performance near the upper
band edge has decreased dramatically with respect to
\citet{Sta99}. Nevertheless we have tentatively detected the
H$_2$D$^+$ ground-state line (Fig. \ref{f2}).
 
The HDO $J$=$1_{01}$--$0_{00}$ ground-state transition (464.924520
GHz) was observed in 1998~July at the JCMT during a night of excellent
submillimeter transparency with a zenith optical depth at 225 GHz
below 0.05. The dual channel, dual band 460/660 GHz SIS receiver RxW
was used in SSB mode with the Digital Autocorrelator Spectrometer
(DAS) backend. The DAS was split in two parts of 125 MHz with a
spectral resolution of 189 kHz (=0.12~km~s$^{-1}$ at 465 GHz).  The
HDO line was observed in the upper side band, and SSB system
temperatures including atmospheric losses were $T_{\rm sys}\la 1000$
K. The beam size is about 11$''$ at 465 GHz and the main beam
efficiency was measured to be about 50\%. We observed the HDO line
toward I16293A in position switch mode with the same reference as
above. We also observed the line in beam switch mode, switching over
$\pm 180''$ in Azimuth. These spectra show the same profile as the
position switched ones, but only the beam switched spectra clearly
reveal the level of the continuum emission from the dust heated by the
YSO. All spectra were coadded and corrected for the continuum
level. In addition, deep spectra of the HDO line toward I16293E were
obtained in position switch mode. The data were calibrated similarly
to the RxA2 and A3 observations.

The H$_2$O $1_{10}$--$1_{01}$ ground-state transition (556.936002 GHz)
was observed in 1999~August with SWAS \citep{mel00}. The receiver
consisted of a Schottky diode mixer resulting in a double sideband
system temperature of about 2200 K.  The back-end was an Acousto
Optical Spectrometer with a 1.4 GHz bandwidth and a velocity
resolution of $\le 1$ km~s$^{-1}$. The beam width is about $4'$. The
H$_2$O line was observed in position switch mode, switching to an
emission free reference position 1.5\degree\ away. The total (on+off)
integration time was 23.5 hrs, yielding a noiose level of $T_{\rm
A}^*(rms)=0.01$ K.

The CO $J$=6--5 (691.473076 GHz) and 7--6 (806.651806 GHz) transitions
toward I16293A have been taken with the JCMT using the RxW receiver in
1999~August and with the MPIfR/SRON 800~GHz SIS receiver in
2000~April, respectively, under good atmospheric conditions. The beam
parameters at these frequencies were determined from observations of
Mars. The main beam efficiencies are about 0.3 and 0.24, and the beam
sizes (FWHM) are $8''$ and $6''$ at 691~GHz and 806~GHz, respectively.

\section{Results and Analysis\label{Results}}

\subsection{Submillimeter Continuum: Morphology\label{continuum}}

IRAS 16293$-$2422 was found to be a binary system in 6~cm and 2~cm VLA
observations by \citet{Woo89}. The two components are separated by
5\asec\ (840 AU) with a position angle of $PA=135$\degree.  High
resolution millimeter observations \citep{Mun92, Loo00} also resolve
the IRAS source into a binary system, in good agreement with the VLA
data.  The southern component (A or MM1) is more extended in
millimeter continuum and is associated with dense gas, high velocity
emission and H$_2$O masers \citep{Mun92, Cho99, Loo00} suggesting that
it is the more active component of the binary system. The northern
component, B or MM2, is very compact both at centimeter and millimeter
wavelengths, and has a surprisingly steep spectrum, $\alpha = 2.3 \pm
0.3$ \citep{Mun92}. It is brighter than A at both 1.3~cm and 2.7~mm
\citep{Est91,Mun92,Scho03}.

\begin{figure} % Figure 4
  \epsscale{1.0}
\plotone{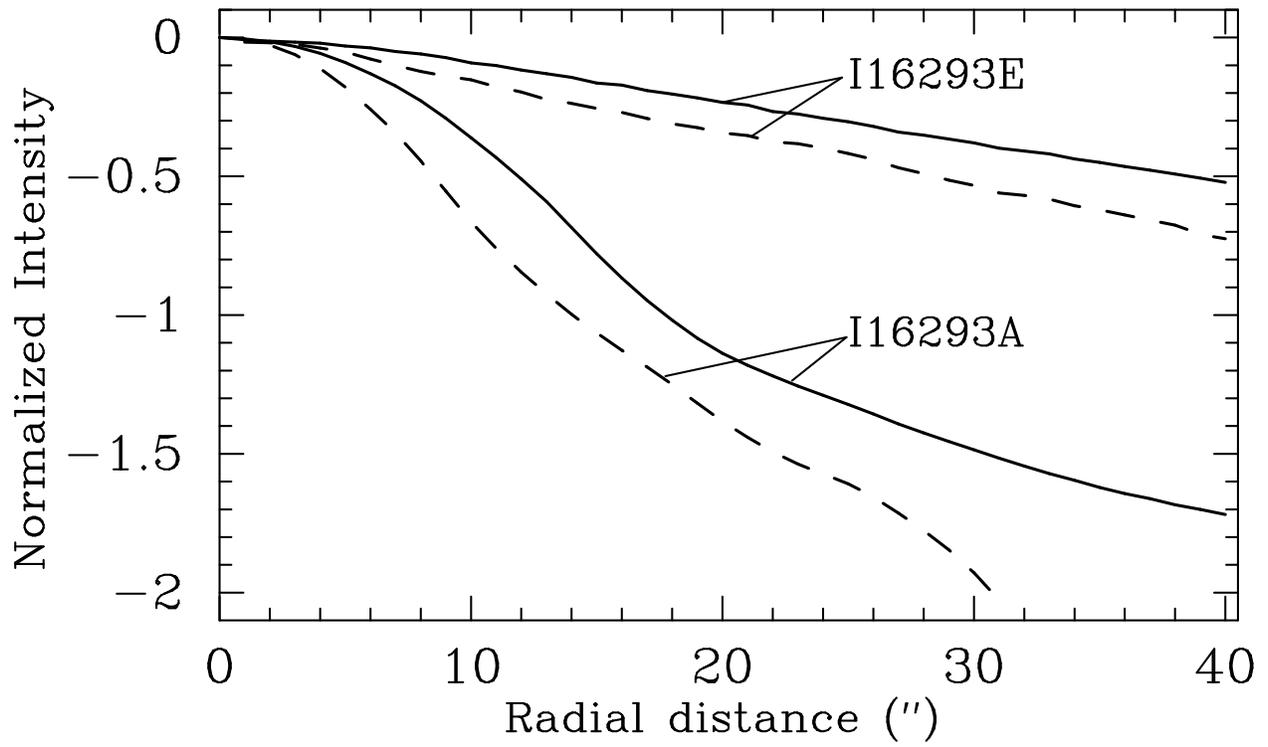}
  \figcaption{Azimuthally averaged flux densities of I16293A and
    I16293E at 850\mic\ ({\sl solid curves}) and 450\mic\ ({\sl dashed
    curves}).\label{f4}}
\end{figure}

We resolve IRAS 16293$-$2422 in all maps with a FWHM $\sim 9.6''
\times 5.7''$ at 450\mic\ with $PA=152$\degree, and find it slightly
more extended at 800 and 850\mic\ with $PA=145$\degree\ and
147\degree, respectively. The corresponding size at 850 $\mu$m is
$\sim 10.0'' \times 7.4''$.  The emission is centered on A to within
1\asec\ in all single dish maps. The same is true for the lower S/N
single-coverage UKT14 maps at 1.1\,mm, 750, 450 and 350\mic. In
Fig. \ref{f4}, we show the azimuthally averaged flux densities of
I16293A and its surrounding envelope, and of the prestellar core
I16293E.  The intensity falls off much more rapidly for I16293A than
for the prestellar core I16293E at radii $r<20$--$30$\asec, beyond
where the slope flattens out to be similar to the prestellar core.

The size and $PA$ seen in our submillimeter maps agree within the
errors with the C$^{18}$O emission mapped by \citet{Mun92} with OVRO
and differs in $PA$ from the alignment of the two submillimeter
sources A and B. In fact, neither the C$^{18}$O emission nor the 450
$\mu$m emission shows any clear enhancement at the B position which is
the stronger continuum source in the wavelength range 1.3~cm to
2.7~mm. At longer wavelengths (850 and 800 $\mu$m) the $PA$ approaches
that of the binary suggesting that at these wavelengths the emission
from B is still significant. However, none of our maps lines up
exactly with the binary, which is what one would expect if the
emission origniated from a circumbinary disk. \citet{Loo00} claim that
their high resolution maps show that the dust emission is aligned with
the binary. However, considering the S/N in their map, one could
equally well interpret their result as a circumstellar disk
surrounding A with a position angle similar to that seen in C$^{18}$O
and 450 $\mu$m, and B as a separate but partly overlapping source.
Their observations show that about 90\% of the dust emission
surrounding A is spatially extended, while B is largely unresolved at
all spatial scales they could measure. \citet{Scho03} observed the
continuum at 1.37~mm with OVRO at $3''$ resolution and find that
I16293A and I16293B both have a disk with diameter $< 250$ AU.

We conclude that the submillimeter dust continuum and most of the
molecular emission are centered on A and have a disk-like
morphology. The dust disk is surrounded by faint extended emission
from the surrounding dark cloud core which falls off rather rapidly
toward the south and southwest. The dark cloud core is more extended
toward the north and northeast ($\sim 70$\asec{}), and in the west a
narrow dust bridge connects it to I16293E (see Fig. \ref{f1}).

\begin{figure} % Figure 5
  \epsscale{1.0}
\plotone{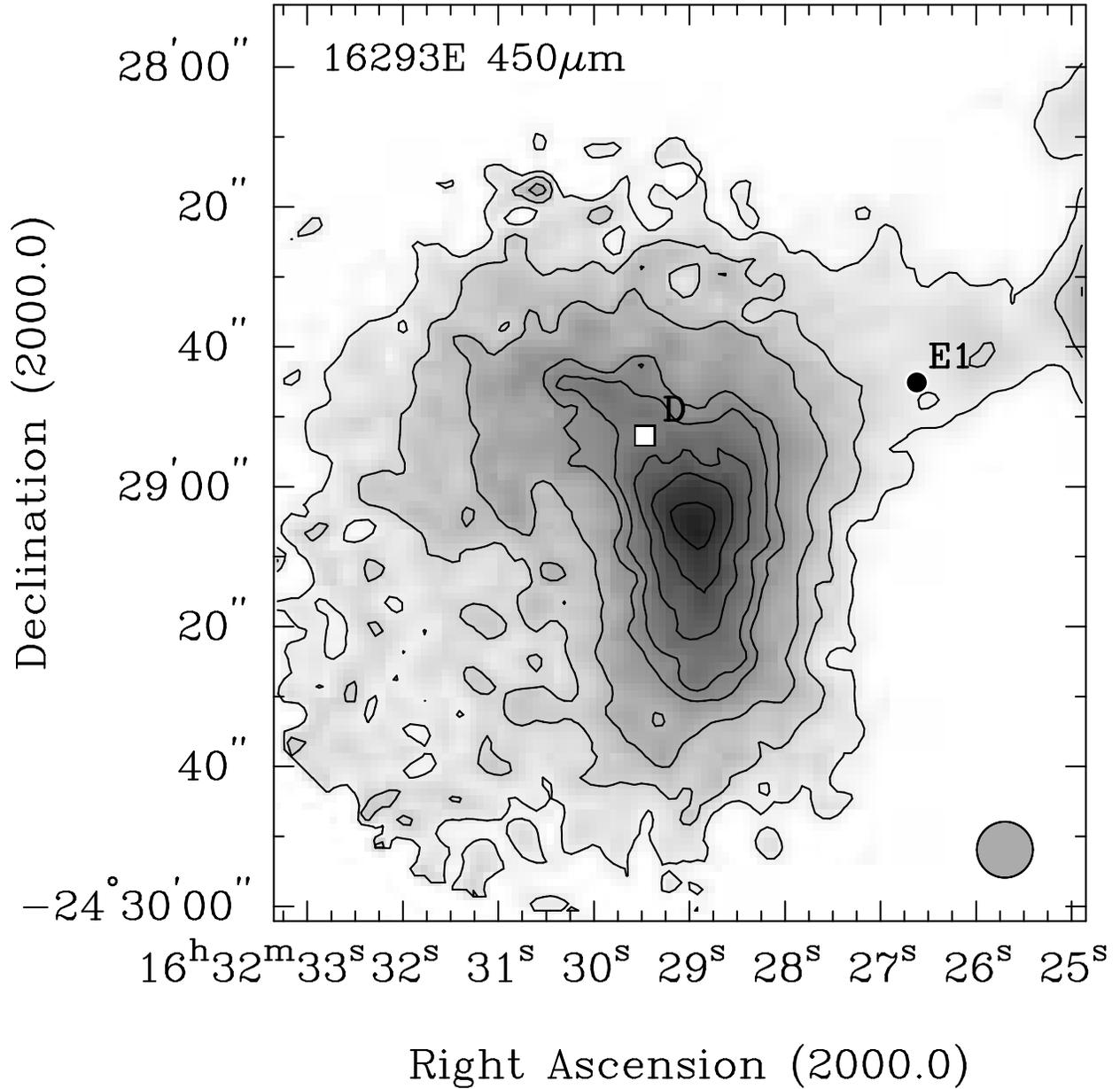}
  \caption{Blow-up of the 450 $\mu$m SCUBA image of I16293E. The
    deuterium peak position D (white square; \citealt{Lis02}) and the
    peak of the SiO emission E1 (black dot; \citealt{Hir01}) are
    indicated.\label{f5}}
\end{figure}

In contrast, I16293E is very extended and has a much flatter emission
profile than I16293A (see Figs. \ref{f1}, \ref{f4}). The peak emission
is centered on \mbox{$\alpha_{2000.0}$ = 16$^h$ 32$^m$ 28.84$^s$},
\mbox{$\delta_{2000.0}$ = $-$24\degree\ 28\amin\ 57.0\asec}\ with a
roughly triangular emission region around the peak. It has a
ridge-like structure at 450 $\mu$m (see Fig. \ref{f5}).  The
integrated fluxes (corrected for error lobe contribution) are 20~Jy
and 133~Jy at 850 and 450 \mic, respectively, in a 50$''$ radius (8000
AU).  In order to perform single position photometry we also made
gaussian fits to our submillimeter images resulting in a source size
of $32'' \times 17''$ with $PA=3\pm 5$\degree\ superposed on a more
extended background.  This source size was used to convert our
photometry into integrated intensities for further analysis.

\subsection{Excitation Analysis}
\label{radial_Tn}

\subsubsection{The Protostellar Disk Around I16293A\label{disk}}

The submillimeter maps allow us to derive a mass estimate for the dust
which is more accurate than previous values. I16293A has been used
extensively as a secondary calibrator for submillimeter continuum
observations at JCMT \citep{San94}, and has accurately known fluxes in
all submillimeter bands.  Although some of our UKT14 maps have
relatively poor S/N, they still allow us to make a reliable estimate
of the flux density of the compact dust disk, because the underlying
emission from the envelope can be determined and subtracted. This is
not possible from photometry data \citep{San94}.  In Table \ref{t1} we
list the integrated fluxes for I16293A from two-dimensional Gaussian
fits to our maps after subtraction of the extended envelope
emission. We have not mapped the region at 1.3~mm, but instead use
data from \citet{Mez92} who derived the integrated flux in a similar
fashion.

We could use the SCUBA data together with co-added IRAS data at 100
and 60 \mic\ to constrain the dust temperature, but it is clear from
our submillimeter maps that the arcmin size of the IRAS beam will also
include emission from the surrounding dark cloud, so to include the
IRAS data we should do a two component fit and simultaneously solve
for the dust envelope and the disk. Instead, we take a simpler
approach. We solve for each separately and use the IRAS data to make
sure that the results we get are plausible. Our deep 850 and 450 \mic\
SCUBA maps predict that about half or more of the continuum emission
seen by IRAS is likely to come from the envelope surrounding the dust
disk. We therefore divide the IRAS 100\mic\ emission into halves and
assume one half to originate in a compact $\leq 7.4$\asec\ disk and
the rest from a cooler cloud envelope around the protostar.

If we assume that the dust grains can be characterized by a single
temperature, $T_{\rm d}$, the flux density $S_\nu$ at frequency $\nu$
can be written as
\begin{equation}                    % EQUATION 2: full blown Planck function
S_\nu = \Omega_s  B_\nu(T_{\rm d})  (1 - e^{-\tau_\nu}) e^{-\tau_{\rm env}},
\label{e2}
\end{equation}
where $\Omega_s$ is the source solid angle, $B_\nu(T_{\rm d})$ is the
Planck function, $\tau_\nu$ the optical depth of the disk, and
$\tau_{\rm env}$ optical depth of the envelope. We assume that the
envelope is optically thin at submillimeter wavelengths ($\tau_{\rm
env}\ll 1$). We write $\tau_\nu$ as
\begin{equation}               % EQUATION 3: definition of optical depth tau
\tau_\nu = \tau_0  \biggl({\nu \over \nu_0}\biggr)^\beta,
\label{e3}
\end{equation}
where $\tau_0$ is the dust optical depth at frequency $\nu_0$, and
$\beta$ is the dust emissivity describing how the dust opacity
($\kappa_\nu \propto \tau_\nu$) changes with frequency \citep{Hil83}.
We adopt the \citet {Hil83} opacity ($\kappa_{1.2~{\rm THz}}$ = 0.1
cm$^2$ g$^{-1}$) and assume a gas-to-dust ratio of 100. It is now
straight forward to do a least-squares fit to equation
(\ref{e2}). Since we mapped the whole cloud, we know $\Omega_s$, and
use this to constrain the fit.  We omit flux densities measured at
3~mm with aperture synthesis telescopes, because these resolve out the
extended emission and may also include free-free emission.

\begin{figure} % Figure 6
  \epsscale{1.0}
\plotone{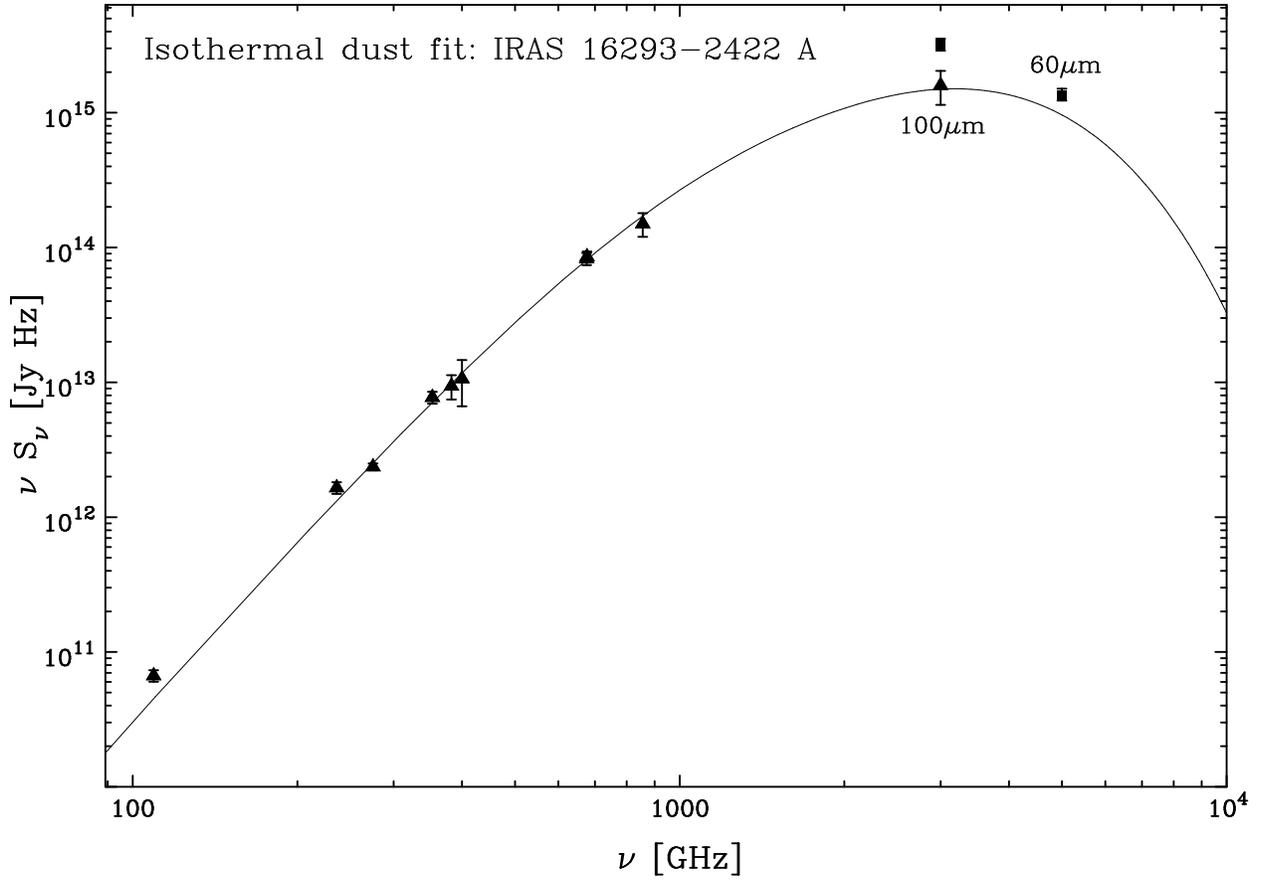}
  \figcaption{Isothermal fit to the continuum emission from
    I16293A. For the IRAS 100\mic\ data we show the total flux, as
    well as the flux estimated to come from the dust disk alone. The
    2.75\,mm (109 GHz) point from \citet{Mun90} is also
    shown. This data point also includes free-free emission from both
    A and B and was not used in the fit.\label{f6}}
\end{figure}

For the submillimeter disk of I16293A we derive a dust temperature
$T_{\rm d}=40 \pm\ 1$ K (see Fig. \ref{f6}), 
$\beta$=1.6 and a source size of about 5\asec. In order to make sure
that these values are not determined by our assumed partitioning of
the IRAS 100 \mic\ flux estimate, we also repeated the fit omitting
the 100 \mic\ data and got similar results. Our fit to the dust disk
is not affected by the 100 $\mu$m data because the dust emission in
this source starts to become optically thin in the (sub) millimeter
regime, $\tau_{850~ \mu{\rm m}} \sim$ 0.42, which acts as a constraint
for the dust temperature. This is unusual as dust emission is usually
optically thin at (sub) millimeter wavelengths and does not provide
any constraints on the dust temperature, unless $\beta$ is known. From
the fit we derive a total mass of 1.8\Msun\, corresponding to an
average density of $n({\rm H}_2)\ge 10^9$ cm$^{-3}$, and a bolometric
luminosity of 16.5\Lsun. The uncertainty in the mass, due to
uncertainties in the fitted $T_{\rm d}$ and $\beta$, is of the order
of 0.3 \Msun. The continuum measurements of \citet{Scho03} show that
the source size is smaller than 3\asec. \citeauthor{Scho03} assume
optically thin emission at 1.37~mm, and a dust temperature of 40~K and
derive a lower limit to the mass of 0.25 \Msun.  Our derived mass of
the dust disk is in the mid-range of the values quoted by
\citet{Mun86}. Our fit appears to slightly overestimate the flux
densities at long wavelengths, which indicates that the data cannot
completely be described by a single dust temperature.

\subsubsection{The Prestellar Core I16293E\label{model_e}}

Since the continuum emission traces the integrated density along the
line of sight, the SCUBA images can be used to derive the density
distribution.  We used the azimuthally averaged radial flux density of
the continuum maps at 450 and 850 $\mu$m to determine the radial
density structure of I16293E.  Figure \ref{f4} shows the radial
emission profiles of this prestellar core; it is clearly seen that
both emission profiles are well described by a power-law.  Instead of
the dust density we will use the H$_2$ volume density as parameter in
determining the density distribution. We adopt a dust-to-gas ratio of
$1:100$ and a power-law H$_2$ density distribution of the form
$n(r)=n_0 (r/1000~{\rm AU})^{-p}$, where $n_0$ is the H$_2$ density at
the arbitrary chosen radius of 1000 AU. Similarly, we adopt a dust
temperature $T_{\rm d}$ distribution following a radial power-law with
index $q$ and temperature $T_0$ at $r=1000$ AU. The dust emissivity is
assumed to be of the form equation (\ref{e3}).

The free parameter set of this model ($n_0,~p,~T_0,~q,~\beta$) is
constrained by the total submillimeter flux, the radial emission
profiles, and the spectral index between 450 and 850 $\mu$m,
respectively.  A $\chi^2$ minimalization between the core model
emission and the data yields best-fit parameters $p=1.0\pm 0.2 $ for
both the 450 $\mu$m and 850 $\mu$m images, $n_0=1.6 \times 10^6$
cm$^{-3}$, and $\beta=2.0$.  The outer radius of the core was set to
8000 AU, the only value of $R_{\rm out}$ which allows a density
profile fit with a single power-law index $p$.  The above density
structure yields a core mass $M_{\rm core}({\rm H}_2)=4.35$ \Msun.
From these results we infer an isothermal dust temperature $T_{\rm
d}=16$ K, i.e. $q=0$.  We apply this temperature to the region $r<
1000$ AU while for $1000~{\rm AU} <r< 8000~{\rm AU}$ we let the
temperature gradually rise to 20 K to include the transition of the
dark core to the diffuse ISM where the temperature is determined by
the interstellar radiation field.  This temperature structure gives a
good fit to the observed radial emission profiles. Note that the
continuum emission is a convolution of the density and temperature, so
the derived density and temperature structures are degenerate, e.g., a
constant temperature $T_{\rm d}=25$ K and $n_0=1.1 \times 10^6$
cm$^{-3}$ also fits the emission profiles.  However, the low-end
$T_{\rm d}$ and high-end $n$ fit yields a plausible physical
description to this prestellar core and best fit to the observed HDO
emision (see below).

The IRAS 100 $\mu$m data were not used to further constrain the above
parameter-set, IRAS was not sensitive to continuum emission from dust
with $T_{\rm d} < 18$ K. On the other hand the SCUBA emission reflects
a convolution of density and temperature along the
line-of-sight. Thus, a cold high density region with $T_{\rm d}\ll 18$
K can easily be hidden in the center of the core since a less dense
warmer envelope component will always dominate the observed continuum
emission.  In general such a region will be hard to detect in
continuum emission as well as in line emission. In the latter case one
would need a transition of a species with a high critical density and
low excitation temperature which is uniquely tracing the coldest and
densest part at the heart of the core and is not depleted.  Deuterated
molecules, and in particular the ground-state transitions of HDO, are
excellently suited to trace such regions (see \S \ref{s:intro} and
below).

The inferred radial density and temperature structures from the dust
are used to calculate the abundance profiles for the observed single
dish observations of C$^{17}$O, C$^{18}$O, DCO$^+$, HCO$^+$,
H$^{13}$CO$^+$, and HDO. The modeling of the molecular excitation and
line radiative transfer uses a spherically symmetric Monte Carlo
method \citep{Hog00b}. The core was divided in 40 concentric
shells. All shells were found to be optically thin at the modeled
transitions of the molecules.  We assume $T_{\rm kin}=T_{\rm d}$
throughout the core and adopt for each line a constant local turbulent
linewidth. After convergence of the level populations was reached, the
spectral line profile of the observed transition of each molecule was
calculated for the appropriate beam. The calculations were done
iteratively starting with an educated guess for the abundance and the
turbulent linewidth, until a best fit of the modeled line profiles to
the observed spectra was established.

\begin{figure} % Figure 7
  \epsscale{1.0}
\plotone{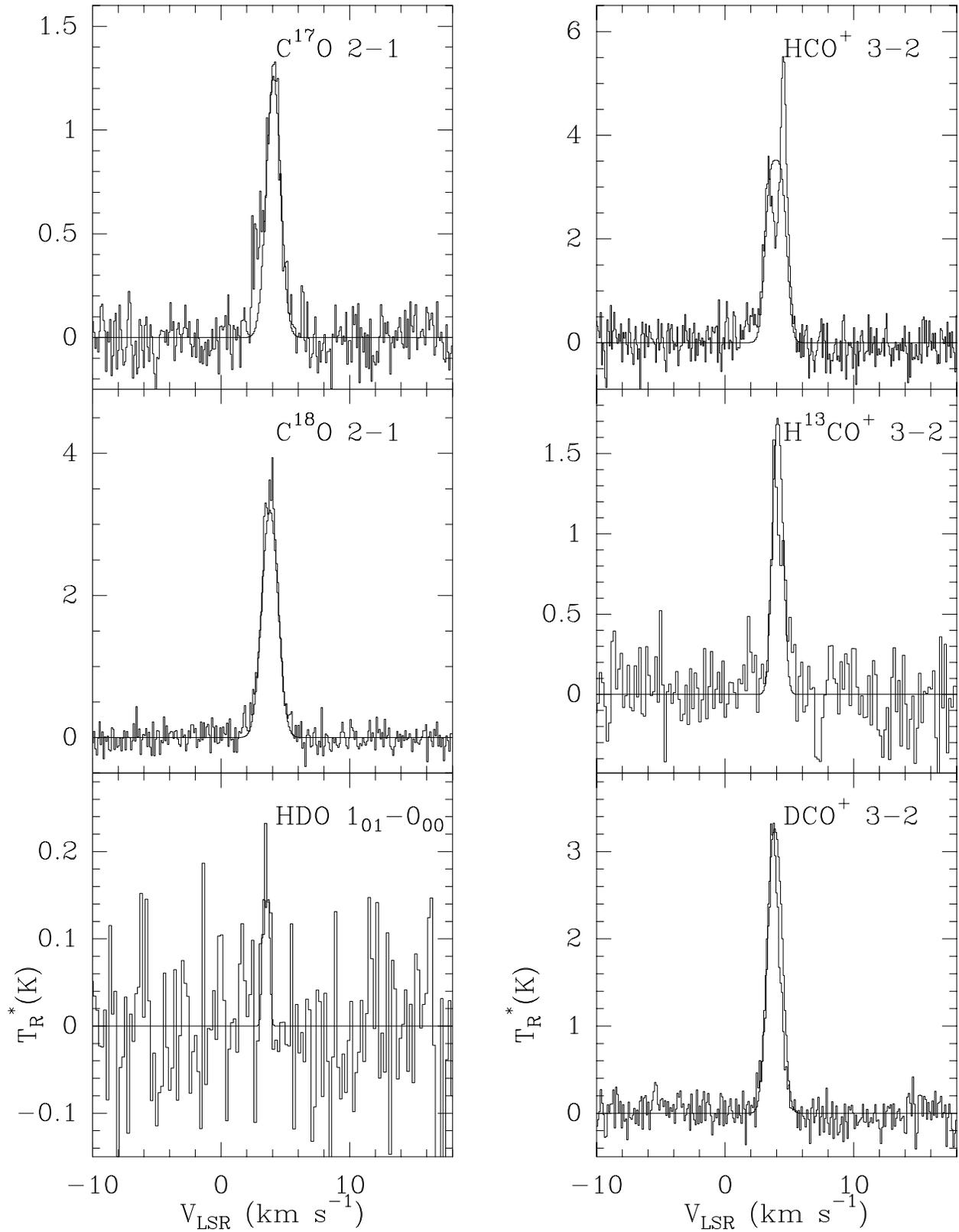}
  \figcaption{Observed spectra toward I16293E overlaid with best-fit
    line profiles from an isothermal core model with a power-law
    density distribution.\label{f7}}
\end{figure}

Our model spectra fit the observed emission profiles for all relevant
molecules (Fig. \ref{f7}) and yield the following abundances for the
CO-isotopomers: [C$^{18}$O]/[H$_2$]=$3 \times 10^{-8}$ and
[C$^{17}$O]/[H$_2$]=$ 9 \times 10^{-9}$.  Adopting standard CO
isotopomer ratios $[{\rm CO}]:[\rm{C}^{18}{\rm O}]=500:1$, and
$[\rm{CO}]:[{\rm C}^{17}{\rm O}]=2500:1$ \citep{Wil94}, our derived
abundances imply that CO is depleted by a factor of about ten with
respect to the standard abundance.

The width and strength of the observed HDO ground-state emission line
is well fit by a [HDO]/[H$_2$] abundance ratio of $2 \times
10^{-10}$. Note that the derived turbulent linewidth of the HDO line
is much lower than that of the other lines.  It is expected that the
turbulent line width is roughly the same for all molecular species
when observed with similar angular resolution if they are distributed
similarly throughout the core.  This is in fact the case for the other
molecules observed with beams of about 15$''$ (DCO$^+$, HCO$^+$, etc.)
for which we obtain an average value of $0.7\pm 0.1$
km~s$^{-1}$. Although the weak HDO emission has a S/N of only 3, its
narrowness suggests that it resides in the innermost region of the
I16293E core. In this cold region ($T_{\rm K}\le 16$ K) the abundance
of a deuterated species like HDO is expected to be enhanced through
equation (\ref{e1}), while non-deuterated molecules are expected to be
depleted (\S \ref{s:intro}). A higher S/N HDO spectrum is required to
confirm this.

The best fit to the HCO$^+$ 3--2 line yields an abundance
[HCO$^+$]/[H$_2$] of $1 \times 10^{-10}$.  The H$^{13}$CO$^+$ 3--2
emission can be reproduced for a constant abundance of $2 \times
10^{-11}$ throughout the core, and the DCO$^+$ 3--2 line fit yields an
abundance of $5\times 10^{-11}$. Note that
[DCO$^+]/[$H$^{13}$CO$^+$]=2.5, that is the DCO$^+$ abudance is larger
than that of H$^{13}$CO$^+$.  Assuming [CO]:[$^{13}$CO]=65:1
\citep{Wil94}, we find that HCO$^+$ is depleted by a factor of about
10. Table \ref{t3} summarizes the inferred abundances and local
turbulent linewidths.

\subsubsection{The Envelope Around I16293A\label{envelope}}

In this section we determine a radial density and temperature
structure for the YSO envelope using an inside-out collapse power-law
density structure where the slope of the density and the velocity
depend on the location of the collapse expansion wave.

We assume that the dust temperature follows a radial power law. Such a
distribution is expected for a spherical cloud with an embedded
heating source in its center. Together with the FIR luminosity and
distance of 160~pc, the temperature distribution is constrained from
the total observed submillimeter continuum emission from the disk and
the envelope (Table \ref{t1}). The inferred temperature profile ranges
from $T_{\rm d}=115$ K at the inner radius $r_{\rm in}=100$ AU, to
$T_{\rm d}=12$ K at the outer radius $r_{\rm out}=7300$ AU.

We use the spherically symmetric self-similar solution of a collapsing
cloud core derived by \citet{Shu77} to determine the density
profile. This model is characterised by only two parameters: the sound
speed $a$, and the time $t$ since the collapse starts at the center at
$t=0$. The initial stationary density distribution is an isothermal
sphere where the density is proportional to $r^{-2}$. At a time $t$
after the onset of collapse, the head of the collapse expansion wave
is radius $r_{\rm CEW}=at$. Inside this radius the velocity field
varies from stationary to free fall, $V\propto r^{-\case{1}{2}}$,
while the density varies as $n\propto r^{-\case{3}{2}}$.  We determine
the infall parameters independently from the molecular spectroscopic
data as well as from the dust continuum measurements. We start with
the former.

We do not use RxA2 observations, which instrumentally broadens the
lines, but only 345~GHz window spectra to determine the infall
parameters.  Earlier studies often used emission lines from lower
rotational transitions of HCO$^+$ or H$^{13}$CO$^+$. Our line profiles
of HCO$^+$ 4--3 and H$^{13}$CO$^+$ 4--3 are complex with excess
emission at the red and blue wings of the doubly peaked emission
profile, respectively (see below). This points to a contribution by
the outflow(s).  The red and blue wings of the DCO$^+$ 5--4 spectrum
are symmetric, and we therefore use this line profile to constrain the
parameter set $a$ and $t$. These parameters respectively determine the
width and integrated line intensity of the profile.

\begin{figure} % Figure 8
  \epsscale{0.9}
\plotone{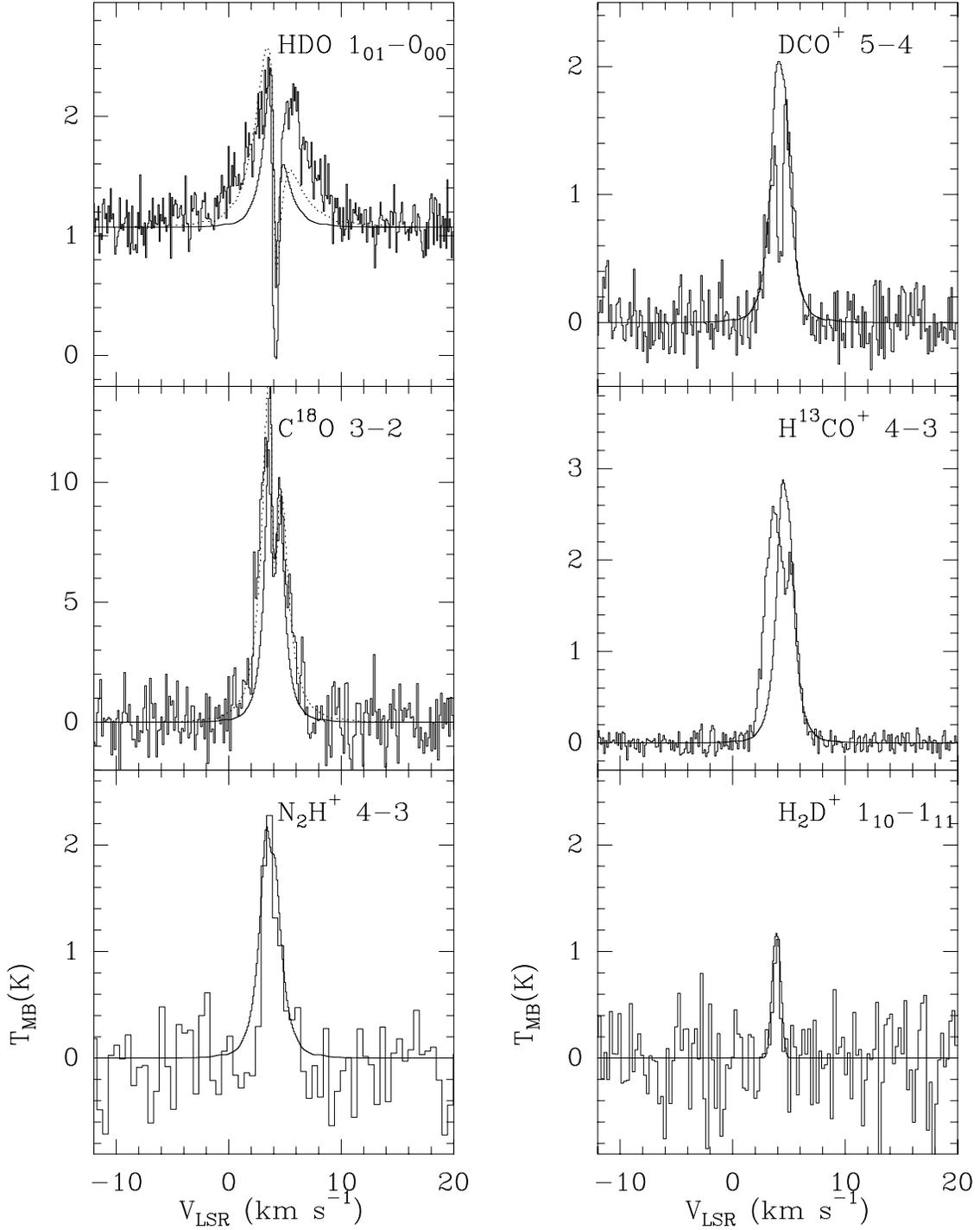}
  \caption{Selection of observed spectra toward I16293A overlaid with
    line profiles predicted by the inside-out collapse model for a
    sound speed $a$=0.7 km~s$^{-1}$ and age $t$=$2.5 \times 10^4$ yr.  A
    better fit to the wings of the C$^{18}$O and HDO spectra is
    obtained for a sound speed $a=0.9$ km s$^{-1}$ and $a=1$ km
    s$^{-1}$, respectively ({\sl dotted curves}).\label{f8}}
\end{figure}

\begin{figure} % Figure 9
  \epsscale{0.7}
\plotone{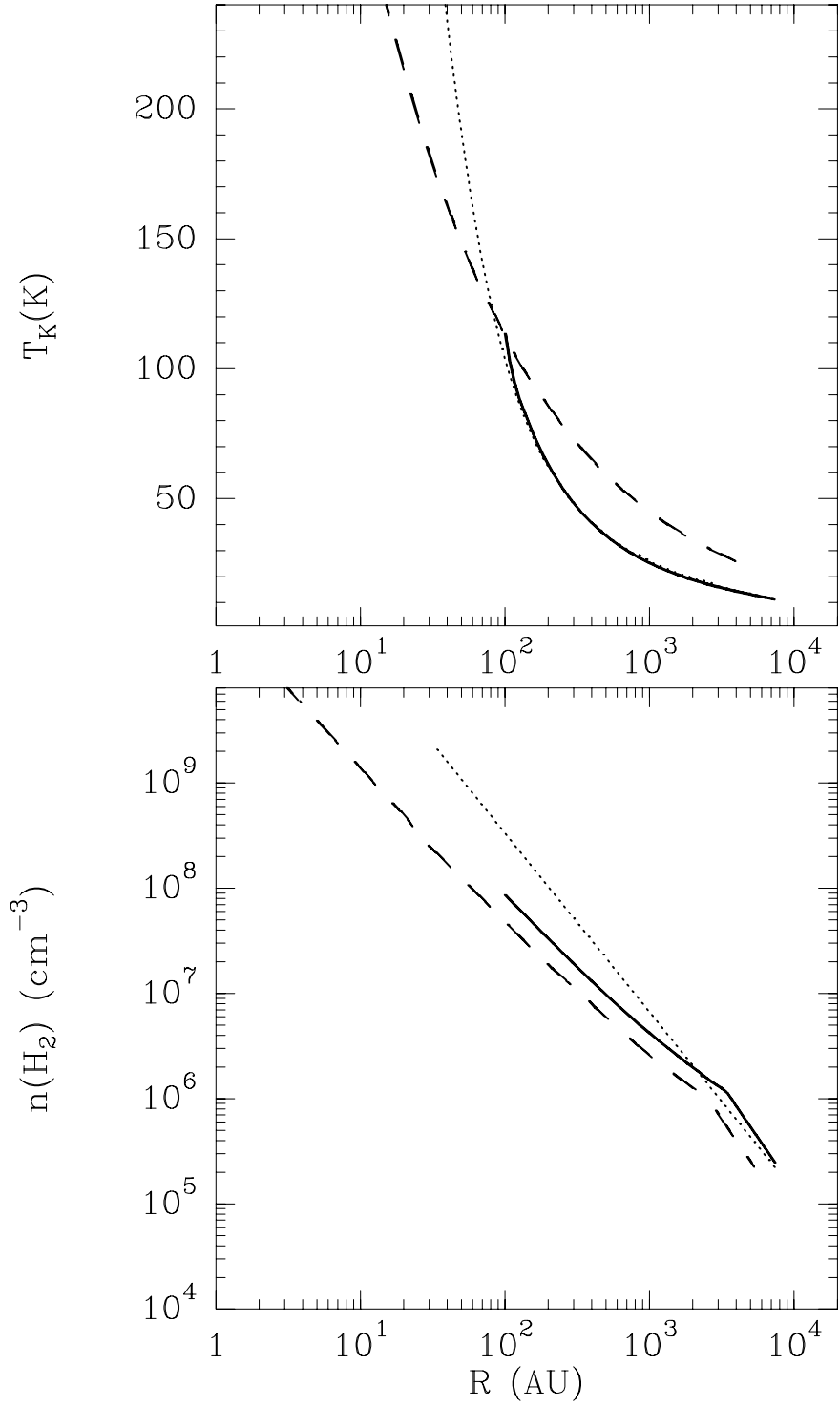}
  \caption{Comparison of our power-law density and temperature
    distributions ({\sl solid curves}) with those of \citet{Scho02}
    ({\sl dotted}) and \citet{cec00} ({\sl dashed curves}).\label{f9}}
\end{figure}

A good fit to the wings of the DCO$^+$ 5--4 emission is found for
$t=2.5 \times 10^4$ yr, $a=0.7$ km s$^{-1}$, and a constant radial
abundance [DCO$^+$]/[H$_2$]=$ 2 \times 10^{-11}$ (Fig. \ref{f8}).
This yields an envelope H$_2$ mass $M=6.1$ $M_\sun$ within a radius of
7300 AU. Note that the absorption can easily be fitted by adding a
cold outer shell where the temperature has dropped to $T_{\rm d}< 10$
K.  Our derived stationary envelope mass corresponds well with the
mass derived by \citet{Scho02} of $M=5.4~M_\sun$ inside $r_{\rm
out}=8000$ AU, but is as much as a factor of four higher than the
masses derived by \citet{Cec96} ($M=1.88~M_\sun$, $r_{\rm out}=5307$
AU) and \citet{Nar98} ($M=2.3~M_\sun$, $r_{\rm out}=6000$ AU).  In
Fig. \ref{f9} we plot the calculated dust and density profiles.  For
comparison we also plot the independently derived density and
temperature profiles from \citet{Scho02} and \citet{Cec96} on a
log-log scale, so power-law distributions become straight lines and
the outer regions of the envelope stand out more clearly. Our profiles
agree well with those of \citet{Scho02} who derived the density and
temperature profiles independently from a subset of our continuum
measurements.  Our values for the infall parameters compare well with
those derived by \citeauthor{Scho02} ($a=0.65-0.95$ km s$^{-1}$,
$t=(1.5$--$3.5) \times 10^4$ yr) and \citeauthor{Cec96} $(a=0.5,~t=2.3
\times 10^4$ yr). The age is more than a factor of two lower than
derived by \citet{Nar98}.  This is probably because our study is
focused on high-$J$ transitions, and therefore more sensitive to the
warm inner envelope regions, while previous work dealt mostly with
lower $J$ transitions which are dominated by the gas in the colder
outer envelope regions.

We used the azimuthally averaged radial emission profile
(Fig. \ref{f4}) to determine independently the infall parameters from
the SCUBA continuum maps. A best fit of the inside-out collapse model
from \citet{Shu77} yields a sound speed $a=0.7$ km s$^{-1}$ and an age
$t=(0.6$--$2) \times 10^4$ yr, close to the values derived from our
fits to the line profiles.  This indicates that the model of
\citeauthor{Shu77} can be used to predict the infall velocities for
this class of YSOs, but that the age is better constrained from the
line profiles. \citet{Scho02} also found a discrepancy between the
infall parameters derived from dust continuum and molecular line
emission.  The infall analysis of \citet{Hog00a} for a sample of YSO
envelopes indicate that this discrepancy may be a common property of
Class~0 YSOs, since a better correspondence was found for Class~I and
older sources.  A possible explanation is that collapse models
generally predict a much higher mass-accretion rate early in the
evolutionary stage than the \citeauthor{Shu77} model does.  We
therefore use the infall parameters $a$ and $t$ from the DCO$^+$ fit
to model the profiles of the observed molecular lines and to infer the
abundances, the results are presented below.

\subsubsubsection{HDO\label{hdo}}

The best fit to the observed HDO profile implies a constant abundance
of $3 \times 10^{-10}$ throughout the envelope. The model profile is
narrower than observed. This may indicate a significant contribution
to HDO emission from, e.g., the outflow(s) (\S \ref{outflow_model}).
To reproduce the line wings by infall, the sound speed must increase
to $a=1$ km~s$^{-1}$.  The fit to the absorption could be improved by
adding a cold shell with $T< 10$ K.  But, HDO collision rates are only
available for $T\ge 50$ K \citep{Gre89}. Our models use the 50~K rates
at lower temperatures while the true rates may easily be a factor of
two lower at $T\simeq $12--15~K enhancing the optical depth of the HDO
line. We therefore expect that extended collision rates would improve
the fit of the line profile.  The continuum level results from the
included FIR radiation field in the excitation analysis and matches
the observed level very well (Fig. \ref{f8}).

We use the best HDO infall model fits ($a=1$ km s$^{-1}$, $t=2.5
\times 10^4$ yr) to model the HDO $2_{11}$--$2_{12}$ and
$3_{12}$--$2_{12}$ lines. These lines have energy levels far above
ground and are not expected to be collisionally excited for this
temperature range. However, in high mass YSOs these lines are found to
be much stronger than expected on basis of collisional excitation
\citep{Jac90, Hel96b, Gen96}, probably because the high levels are
populated through excitation by the FIR radiation field. With the
earlier mentioned dust parameters for the FIR excitation, the model
yields a $2_{11}$--$2_{12}$ peak intensity of $T_{\rm R}^*=0.006$ K,
and $3_{12}$--$2_{12}$ $T_{\rm R}^*=0.001$ K.  The former value is
much lower than the upper limit to the line strength of 0.08~K from
\citet{evd95} for the $2_{11}$--$2_{12}$ line. Since the upper-limits
of \citeauthor{evd95} for the $2_{11}$--$2_{12}$ and
$3_{12}$--$2_{12}$ lines are high with respect to what can be achieved
with present-day, more sensitive receivers, we decided to reobserve
both lines.  Deep spectra reveal no features at a noise rms of 0.02~K
at either frequency for a spectral resolution of 1.25 MHz (=1.7
km~s$^{-1}$).

\subsubsubsection{H$_2$O\label{h2o}}

The excitation of H$_2$O has been calculated using the collision rates
of ortho- and para-H$_2$ with ortho-H$_2$O \citep{Phi96} for kinetic
temperatures between 20 and 140~K for the lowest 5 rotational energy
levels of H$_2$O. This range traces the temperature structure of our
model very well. We interpolated the rates for the appropriate
temperature of each shell.  The ortho/para H$_2$ ratio in each shell
is assumed to be in LTE at the appropriate temperature of each shell;
for low temperatures, $T_{\rm kin}< 70$ K, most of the H$_2$ is in its
para modification and [ortho-H$_2$]/[para-H$_2$]$ \rightarrow 0$ for
lower $T_{\rm kin}$.

We included FIR dust excitation in our model, using the dust
emissivities from \citet{Oss94} . Calculations with a large range in
power-law emissivities show that the dust emission has not much impact
on the excitation of the lowest water lines in the envelope of
I16293A. Therefore, we did not consider the FIR radiation in our
further calculations. The total H$_2$O abundance is derived assuming
an ortho-to-para H$_2$O ratio of $3:1$.

The abundance was varied to fit the self-reversal and to match the
$1_{10}$--$1_{01}$ peak H$_2$O emission observed by SWAS.  Our infall
models fail to match the large width of the observed H$_2$O
spectrum. This is not surprising; in the large SWAS beam (FWHM$\simeq
4$ arcmin $\simeq 35,000$ AU) emission from our model cloud is
dominated by the cold outer envelope.  The width of the observed line
profile clearly indicates that most of the H$_2$O emission arises in
molecular outflows associated with this Class~0 YSO which are not
included in our envelope model. We therefore focus here on the region
causing the absorption and the emission at the systemic velocity of
the cloud, and model the H$_2$O outflow in \S \ref{outflow_model}. Our
envelope model predicts a narrow double-peaked profile which can match
the intensity level of the observed maxima, but not the width
(Fig. \ref{f10}). Initially, the analysis was done for a uniform
abundance [ortho-H$_2$O]/[H$_2$]=$2 \times 10^{-7}$ across the
envelope. This fits the maximum emission, but causes too deep an
absorption. To reduce the absorption we applied a step function for
the abundance profile with a drop in the H$_2$O abundance at $T_{\rm
kin}=14$ K, where atomic oxygen starts to freeze out. Such a two-step
abundance profile fits the observations for [ortho-H$_2$O]/[H$_2$]=$3
\times 10^{-9}$ for $T_{\rm kin}< 14$ K (Fig. \ref{f10}). The FIR
H$_2$O lines observed with ISO \citep{cec00} are much stronger than
predicted by our model, indicating that they do not originate in the
envelope.

\subsubsubsection{H$_2$D$^+$\label{h2d+}}

The H$_2$D$^+$ $1_{10}$--$1_{11}$ line has been marginally detected at
a 3-sigma level. A constant abundance of $3\times 10^{-10}$ yields a
good fit to the observed maximum emission, but the line width of the
model spectrum is too large. We therefore tried a step function for
the abundance, [H$_2$D$^+$]/[H$_2$]=$1 \times 10^{-12}$ for $T_{\rm
kin}>$ 20 K and [H$_2$D$^+$]/[H$_2$]=$2 \times 10^{-9}$ for $T_{\rm
kin} < 20$ K which yields a good fit to the complete line profile
(Fig. \ref{f8}).

\subsubsubsection{N$_2$H$^+$\label{n2h+}}

The N$_2$H$^+$ 4--3 line was observed simultaneously with the
H$_2$D$^+$ $1_{10}$--$1_{11}$ line. A constant abundance with the
radius of $3 \times 10^{-11}$ yields a good fit to the observed line
profile (Fig. \ref{f8}).

\subsubsubsection{HCO$^+$\label{hco+}}

We can only put a lower limit to the HCO$^+$ abundance of $1\times
10^{-9}$, where the line 4--3 starts to become optically
thick. However, this does not match the peak intensities well, nor
does it fit the apparent extra blue and red components. The wings are
well matched for $a=1$ km~s$^{-1}$. An outflow is likely contributing
to the emission. The abundance for $T_{\rm kin}<$ 10 K could be
enhanced to get a deeper absorption, but this would not add much
information since the emission and the absorption are already
optically thick. Note that the high optical depth of the emission and
absorption masks any signature of chemical changes at low
temperatures. This transition is therefore not suited for an accurate
derivation of the abundance structure.

\subsubsubsection{H$^{13}$CO$^+$\label{h13co+}}

For this HCO$^+$-isotopomer, we also have problems in modeling the
observed double line structure, probably because kinematically
distinct components are present at the blue side (Fig. \ref{f8}).  An
abundance of $2\times 10^{-11}$ gives a good fit to the red
component. This component is optically thin and we attribute it to the
envelope.  Note that a constant abundance thoughout the envelope is
used for this fit. The blue component is probably associated with one
of the quadrupole outflow components (\S \ref{CO_outflows}).

\subsubsubsection{$^{12}$CO\label{12co}}

The CO 6--5 and 7--6 lines have been modeled for a $^{12}$CO abundance
[$^{12}$CO]/[H$_2$]=$1 \times 10^{-4}$. In Fig. \ref{f11} the model
spectra are overlaid on the observations. It can be seen that the
envelope model fails to fit the observed emission for a sound speed
$a=0.7$ km~s$^{-1}$. Even for $a=1$ km~s$^{-1}$, the width and maxima
of the observed spectra cannot be matched. This indicates that, like
the H$_2$O emission, most of the observed high-$J$ $^{12}$CO emission
is associated with the outflows (\S \ref{outflow_model}).

\subsubsubsection{C$^{18}$O\label{c18o}}

The C$^{18}$O 3--2 line profile can well be fitted for an abundance of
$2\times 10^{-7}$ throughout the envelope, corresponding to a
`standard' [CO]/[H$_2$] abundance of about $10^{-4}$.  \citet{Scho02}
derive a C$^{18}$O abundance of $6.2\times 10^{-8}$ on basis of
similar observations with the JCMT. However their abundance yields a
fit to the spectrum which is about a factor of two lower than our fit
(see their Fig. 4), so the agreement is within a factor of 1.5.  Our
abundance-fit yields a C$^{18}$O 2--1 spectrum whose central region is
quite different from the observations indicating a lower abundance.
The observed low emission level of the C$^{18}$O 2--1 may be caused by
a low-density C$^{18}$O component with a `standard' abundance residing
in dark/translucent transition region between the outer envelope and
the diffuse interstellar medium, while CO in the cold outer envelope
regions may be highly depleted. Moreover, the modeled C$^{18}$O 2--1
and 3--2 line profiles are somewhat narrower than observed.  The line
wings of both transitions can be fitted by infall if a sound speed of
$a=0.9$ km~s$^{-1}$ is adopted, or there may be a contribution from
the outflow.  Figure \ref{f8} plots the excitation model fit for the
C$^{18}$O 3--2 line.  We conclude by noting that the low-$J$ C$^{18}$O
transitions are not well suited to determine the CO depletion in the
envelope since the line profile indicates than C$^{18}$O is sensitive
to material which may not reside in the envelope.

\subsubsubsection{C$^{17}$O\label{c17o}}

We use a constant C$^{17}$O abundance of $4\times 10^{-8}$, i.e., a
`standard' CO abundance for $T_{\rm kin}> 20$ K and depletion by a
factor of 50 for $T_{\rm kin} < 20$ K.  This fits better than a
uniform abundance throughout the envelope, because depletion makes the
line appear broader. The line is also broadened by unresolved
hyperfine lines, e.g., visible as a low-velocity shoulder. Still, the
observed line is broader than our model predicts. The wings fit better
for a sound speed $a=0.9$ km~s$^{-1}$. There is a hint for a
depression of the peak emission. \citet{Scho02} derive a C$^{17}$O
abundance of about $2\times 10^{-8}$ which agrees within a factor of
two with our results using a constant abundance value.

\subsubsubsection{H$_3$O$^+$\label{h3o}}

The ortho-H$_3$O$^+$ $3^+_0$--$2^-_0$ line at 396.272412 GHz has been
searched for in I16293A by \citet{Phi92}. We used their upper limit
and our envelope excitation model to constrain the abundance
[ortho-H$_3$O$^+]/[$H$_2$]$<1.5 \times 10^{-9}$. This correspondes to
an [H$_3$O$^+]/[$H$_2$] upper limit of (3--5)$ \times 10^{-9}$ if the
H$_3$O$^+$ resides mainly in the $T_{\rm kin}<50$ K region, or $T_{\rm
kin}>50$ K region, where the [ortho-H$_3$O$^+$]:[para-H$_3$O$^+$]
ratio varies between $1:2$ and $1:1$, respectively. These upper limits
are about an order of magnitude larger than derived by \citet{Phi92}
using statistical equilibrium calculations for a slab with a
homogeneous density and temperature.

The derived abundances of all species are summarized in Table \ref{t3}
together with those of \citet{Scho02}. Taking into account the
differences between their fits to the observed spectra and ours (see
above), the abundances agree within a factor of 1.5.

\section{The CO Outflows and What Drives Them}
\label{CO_outflows}

\subsection{The NE--SW Flow: Powered by I16293A\label{nesw}}

The symmetry axis of the NE--SW outflow, as determined from our CO
2--1 map, goes through A at all velocities (Fig. \ref{f12}). The
centroid of the overlapping blue- and red-shifted emission is centered
on I16293A, which is also evident from the velocity position plot in
Fig. \ref{f13} (although inescapably includes some of the E--W
outflow). Furthermore, aperture synthesis maps of dense outflow
tracers, especially CS \citep{Walk93} and SO \citep{Mun92} clearly
show high velocity gas emanating from I16293A in the direction of the
CO outflow, while I16293B shows no activity, nor does it seem to be
associated with a maximum in any high density gas.  We therefore
attribute the NE--SW outflow to I16293A. Our submillimeter maps
(Figs. \ref{f1} and \ref{f4}) clearly show that the dust emission
associated with the free-free and millimeter-source I16293A is
extended. This is one of the few YSOs where we can resolve the dust
emission: it has a disk-like structure with a size of roughly 2000~AU.
At low frequencies the continuum emission from I16293A splits up into
a double source. We do not believe that I16293A is in itself a binary;
a more plausible interpretion is that we are seeing highly collimated
ionized gas from the outflow, as has been seen in L1551~IRS5
\citep{Bie85,Rod86}, where the jet dominates the emission at low
frequencies and the disk emission becomes dominant at high
frequencies.  These are strong arguments for an accretion disk that is
centered on I16293A and drives a powerful outflow, which is very well
collimated close to the star (e.g., free-free emission), but
relatively highly collimated even in CO.  The ratio of length to width
is about 1:3 at high velocities. The outflow is at high inclination,
since we can see overlap between the blue- and the red-shifted
lobe. If we assume the outflow to be conical, which according to the
velocity position plot (Fig. \ref{f13}) appears to be a reasonable
assumption, we can use the method outlined by \citet{Lis86} to
estimate the inclination of the outflow, and find about 65\degree\ for
both the blue and the red lobe. This is slighly higher than what
\citet{Hir01} find on the basis of their SiO 2--1 map for the NE red
lobe, where they derive an inclination by 40--45\degree\ from the
plane of the sky.

The properties of the outflow have been derived in several studies
(e.g., \citealt{Walk88, Miz90, Cas01, Hir01, Gar02, Lis02}, but we
will redo it here; partly because we have more accurate data, but also
to point out how uncertain these estimates are. The deep integrations
of $^{13}$CO and C$^{18}$O toward I16293A (Fig. \ref{f2}) clearly show
high velocity emission, which indicates that the $^{12}$CO optical
depth in the wings is substantial. At near outflow velocities ($\sim$
1.5\kms\ from the cloud systemic velocity) we find optical depths of
40 or more, both from $^{13}$CO and C$^{18}$O, but note that here the
line wings originate from both outflows and also get a significant
contribution from the underlying accretion disk. We therefore obtained
additional $^{12}$CO and $^{13}$CO spectra at the peak positions in
the lobes of the NE--SW outflow, i.e., at offset positions
(+80\arcsec,+55\arcsec{}) for the redshifted lobe and
($-$130\arcsec,$-$40\arcsec{}) for the blueshifted lobe. These show
that the red-shifted outflow has substantially higher optical
depth. We estimate a $^{12}$CO 2--1 optical depth of about unity at
radial velocities as high as 11\kms\ away from the systemic velocity
of the cloud, whereas the optical depth has dropped to unity at
+6\kms\ away from the cloud core for gas in the blue-shifted
outflow. The $^{13}$CO spectra are not sensitive enough to enable us
to derive optical depths of less than about unity.

We compute the outflow mass, momentum, momentum flux, and energy
content with and without opacity corrections, by assuming that we can
apply the same opacity correction to all of the gas in the
outflow. Since we have not directly measured the excitation
temperature of the outflowing gas we assume it to be 50~K, but note
that it could be as high as 80~K \citep{evd95}. The values in Table
\ref{t4} are obtained assuming an abundance ratio 10$^{-4}$ for
[CO]/[H$_2$] and summing over the map in 1 \kms\ wide velocity
intervals starting 1 \kms\ from the cloud systemic velocity (assumed
to be 4 \kms{}).  The apparent dynamic outflow time scales are about
the same for both lobes, (3.0--3.5)$\times 10^3$ yr.  This agrees with
the dynamical timescale of (5--7)$\times 10^3$ yr derived by
\citet{Hir01} for the NE red lobe.  However, the blue outflow lobe
appears more energetic.  Normally when one sees an imbalance in the
momentum or force of the outflow, it is an indication that one of the
outflow lobes has penetrated the cloud surface and therefore has no
gas to interact with. Here we see no evidence for this. A more likely
explanation is that we have not properly corrected for the opacity and
excitation differences in the two outflow lobes. The blue lobe is more
extended and has higher apparent velocities, suggesting that the gas
surrounding the blue lobe is less dense than the gas surrounding the
red outflow lobe.  This is also seen in our opacity estimates, which
are higher for the red-shifted outflow. A similar conclusion is
reached by \citet{Hir01} and \citet{Gar02} on basis of SiO and
CH$_3$OH maps, respectively.

\subsection{The E--W Outflow: A Fossil Flow Driven by I16293B\label{ew}}

The symmetry axis of the E--W outflow passes north of I16293A, but not
as clearly defined as the NE--SW outflow. The red-shifted outflow lobe
is much more compact than the blue-shifted lobe, and appears to bend
more towards the north. However, the peak intensity of the outflowing
red-shifted gas is rather well aligned with the large blue-shifted
outflow lobe with an average position angle for the flow of $PA\simeq
96 \pm 3$\degree.  The E--W outflow appears to intersect with the
continuum disk about 5--10\arcsec\ north of I16293A. Since there is no
source near this position other than I16293B, we associate the E--W
outflow with I16293B. A similar conclusion was also drawn by
\citet{Walk93}, although they stress that at present there is no
evidence for outflow activity in the immediate vicinity of
B. \citet{Gar02} find that the CH$_3$OH wing emission has a narrow and
nearly constant velocity width along the E--W lobes, which they
attribute to turbulent entrainment. This is consistent with our
finding that the E--W outflow is a fossilized flow.

It has been suggested that the NE red outflow lobe and the eastern
blue outflow lobe could be part of a wide angle flow similar to the
L~723 outflow \citep{Walk88} or that the two outflows are part of a
precessing or episodic outflow driven by the same star
\citep{Walk88,Walk93,Miz90}. However, our CO map clearly shows that
these are two separate outflows.  Since the symmetry axis of the E--W
outflow does not intersect with I16293A, it is rather unlikely that
this outflow would have been caused by an earlier outflow episode from
I16293A. \citet{Hir01} and \citet{Gar02} also conclude that two
independent bipolar outflows are present on the basis of their SiO and
CH$_3$OH maps.  \citet{Miz90} also discuss a third outflow, because in
their low spatial resolution Nagoya map, the E--W outflow extends more
than 10\arcmin\ from IRAS~16293$-$2422 and their high resolution
Nobeyama map does not show this extended outflow component. Our map
does not extend as far as the Nagoya map, and our spatial resolution
is poorer than that of the Nobeyama CO 1--0 map. However, the CO map
we present here has the advantage of being fully sampled, unlike the
SiO 2--1 emission mapped by \citet{Hir01}, and it has a much higher
resolution (20$''$) than the large-scale SiO 2--1 map of \citet{Gar02}
($57''$). We therefore better resolve the two outflows. One can
clearly see that the outflow continues to the east outside the area we
have mapped, but with rather low velocities. Therefore the extended
blue low-velocity outflow appears simply a continuation of the E--W
flow.

We also clearly see the interaction of this outflow with the I16293E
core, which was already found by \citet{Miz90}.  In our channel maps,
which are overlayed on the 800\mic-continuum emission
(Fig. \ref{f12}), this interaction is clearly visible. It appears
rather gentle: the outflow is enhanced at low velocities and seems to
stream around the core and forms almost a cavity immediately behind
it. That it streams around the core region is supported by the fact
that we see faint red-shifted $^{12}$CO towards the dense core region,
but the red-shifted emission is much stronger in front of the core,
although one can still see it faintly on the backside as well
(Fig. \ref{f12}). Figure \ref{f5} shows a blow-up of the 450 $\mu$m
image of I16293E. It is clear that the outflow has compressed the core
and caused the high density N--S ridge that we see in our continuum
maps and prominently at near cloud velocities in the channel-maps of
Fig. \ref{f12}. Behind the core it appears that the outflow gets
recollimated. There is no indication of interaction between I16293E
and the red-shifted NE outflow lobe from our CO channel maps, nor from
the SiO channel maps \citep{Hir01, Cas01, Gar02}. \citet{Cas01}
attribute the blue and red stream of the E--W outflow around the
I16293E core to the red and blue lobe of an outflow powered by I16293E
causing two shocked regions (HE1 and E1 in their nomenclature) visible
in their H$_2$CO maps. They support their hypothesis by means of
HCO$^+$ 1--0 spectra taken along a slice which goes through I16293E as
well as through the peaks in the H$_2$CO emission, since these spectra
reveal a blue wing at one side of I16293E and a red wing on the
counter side. Although these wings are real, the results are different
when put in context of the entire cloud. Our large maps show what
small maps, which do not fully cover the complete interaction region
of the extended outflows with the LDN~1689N cloud, cannot: local small
scale flows in the vicinity of I16293E are part of large scale
motions. From our CO map it is clear that the apparent SW--NE outflow
and the alignment with I16293E are due to projection effects of pieces
of the E--W outflow from I16293B which streams around the dense
prestellar core I16293E. This view is supported by the study of
\citet{Hir01} who find that the E--W flow has fan-shaped lobes, each
of which containing blue- and red-shifted components. This would imply
that E1 is red-shifted gas in the far side of the eastern blue
lobe. The red-shifted shocked region E1 is thus the interaction region
of the E--W outflow, while the blue-shifted H$_2$CO peaks HE1 and HE2
from \citet{Cas01} are the interaction of the near side of the E--W
blue lobe with dense gas of LDN~1689N located at the southwest and
southeast side of the dense core I16293E. These interaction regions
could also be the edges of I16293E.

\subsection{The High-$J$ CO, HDO, and H$_2$O Emission: Outflow Modeling}
\label{outflow_model}

Section \ref{envelope} found that the envelope model of I16293A can
reproduce the line profiles of most observed molecular transitions,
but not of CO 6--5 and 7--6, HDO, and H$_2$O. Although the absorption
at the cloud systemic velocity can be fit with an envelope model, the
broad emission cannot be reproduced even for sound speeds as high as
$a=1$ km~s$^{-1}$. Here we investigate the origin of the broad
high-$J$ CO, HDO, and H$_2$O wings by means of an outflow model.

We have scaled our CO 7--6 emission to the level of the H$_2$O
emission (Fig. \ref{f14}). It is striking that the two spectra have
such a similar shape despite the fact that the beam sizes differ by a
factor of about 30. The match of the absorption widths is also very
good, and indicates that we may see here the same region with JCMT and
SWAS in absorption.  In addition, the line wings are similar, although
at the blue side there is an extra CO emission component while at the
red side there is an extra H$_2$O emission component present.  But the
extent of the wings is similar for both species. Does this mean that
the wings are from a very small region covered by both beams, or that
the wing component is rather uniformly distributed throughout the
cloud?

\begin{figure} % Figure 10
  \epsscale{0.85}
\plotone{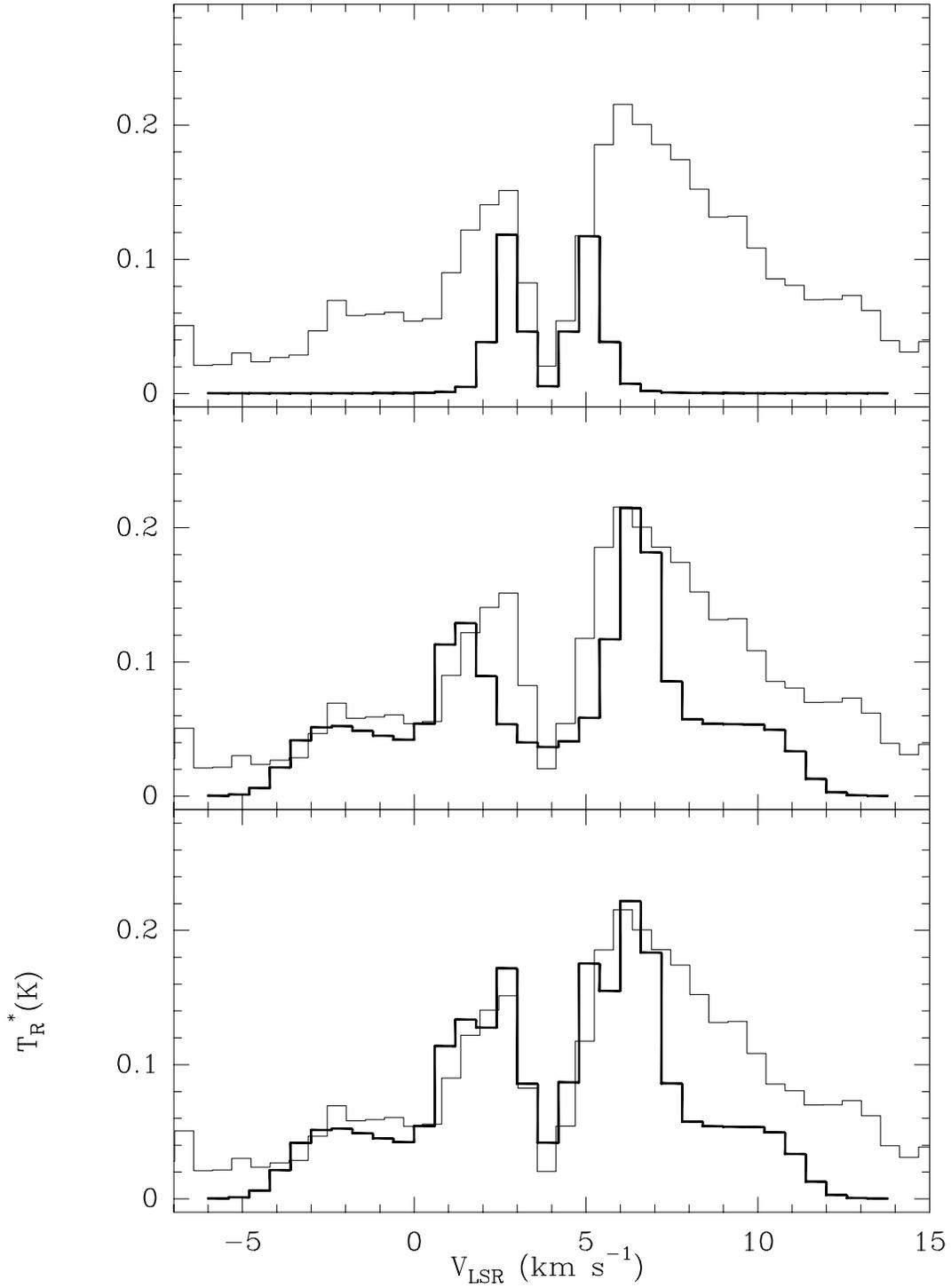}
  \caption{Observed H$_2$O spectrum ({\sl thin curves}) from SWAS
    overlayed with our collapse envelope model spectrum ({\sl top
    panel }); the outflow model spectrum ({\sl middle}); and the
    coadded infall and outflow spectra ({\sl bottom}).\label{f10}}
\end{figure}

\begin{figure} % Figure 11
  \epsscale{0.85}
\plotone{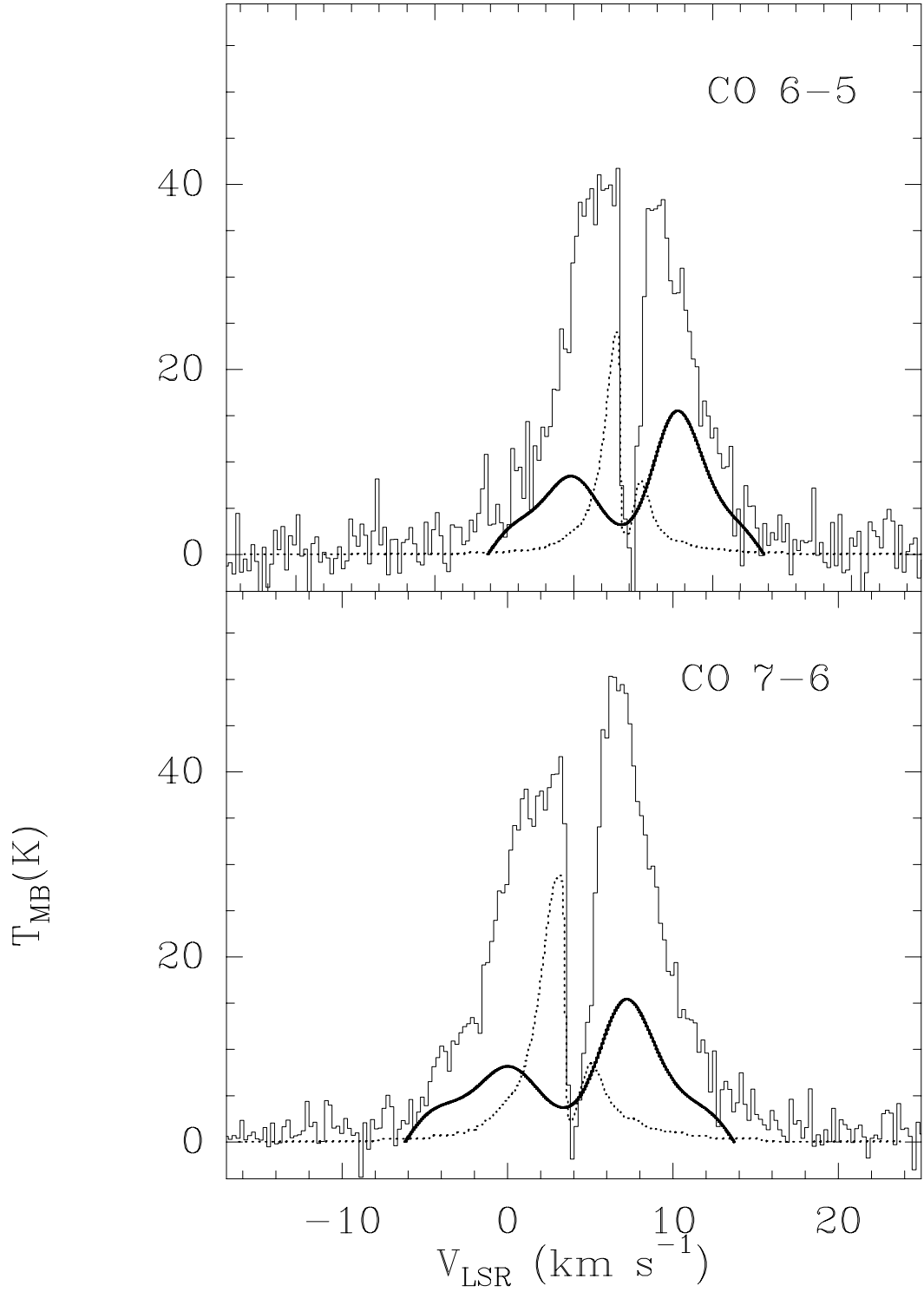}
  \caption{Observed CO $J$=6--5 and 7--6 spectra toward I16293A
    together with overlaid infall ({\sl dotted}) and outflow ({\sl
    thick}) model profiles for a [CO]/[H$_2$] abundance of
    $10^{-4}$.\label{f11}}
\end{figure}

\begin{figure} % Figure 12
  \epsscale{0.40}
\plotone{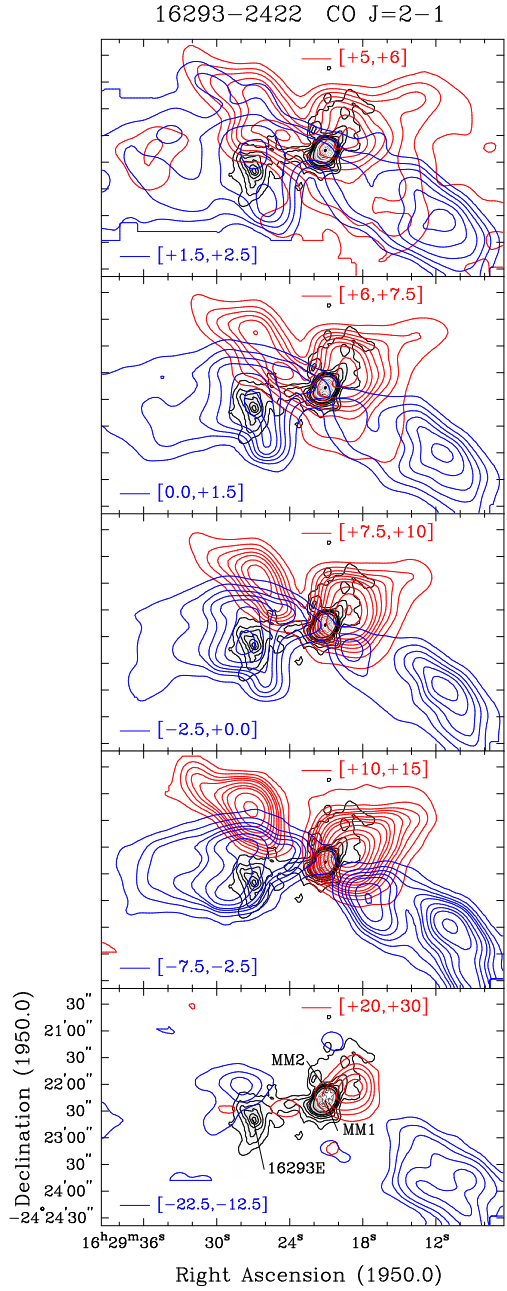}
  \caption{Contour plots of the two outflows from IRAS 16293--2422 A
    and B for indicated velocity intervals superposed on our 800\mic\
    continuum image.\label{f12}}
\end{figure}

The SWAS data were fitted with a Monte Carlo radiative transfer code
and a physical model of a spherically symmetric cloud subdivided into
shells of constant physical conditions. The use of the code and its
application to water lines is described in \citet{Ash00}, which also
analyzes the variety of water line profiles and establishes broad
diagnostics of outflow, infall, and turbulence apparent in SWAS data.
The line shape lead us to explore models which combine outflow and
turbulence; we find that the line strengths depend on the velocity
field as well as on the temperature and abundance in a complex
relation.  Figure \ref{f10} (middle panel) shows the predicted
ortho-H$_2$O spectra from the outflow model with an ortho-H$_2$O
abundance of $1 \times 10^{-8}$ throughout, a turbulent velocity of
2.3 km~s$^{-1}$, and maximum outflow velocity $V=5.8$ km s$^{-1}$. The
temperature ranges from 200 K to 15.5 K.  The outflow model provides
an acceptable fit to the observed extended wing emission, but cannot
explain the central part of the observed profile.  The outflow model
results are complementary to the infall modeling of the ortho-H$_2$O
line (Fig. \ref{f10}, top and middle panel), which fits only the
central region of the observed spectrum. To illustrate this
qualitatively, we plotted in Fig. \ref{f10} (bottom pannel) the sum of
the outflow and envelope spectra of ortho-H$_2$O which indeed yields a
rather good description of the entire SWAS spectrum. Note that the red
wing of the H$_2$O emission is not fitted by our model, this may be
due to multiple outflows (\S \ref{nesw} and \S \ref{ew}) which we did
not model.  For a quantitative comparison we do not only need a
mutiple ouflow model but also have to incorporate the outflow and
infall models in one single excitation model since they overlap in
extent and velocity space. This could mean that part of the blue
infall gets absorbed by the blue outflow component and vice versa. An
avenue for further study would be a combined infall and outflow model.

The same outflow models were used to model the CO 6--5 and 7--6 lines
in the JCMT beam using an abundance [CO]/[H$_2$]=$10^{-4}$
(Fig. \ref{f11}).  The wing emission predicted by the outflow model
poorly fits the observed emission which indicates the presence of
multiple outflows, analogous to the H$_2$O situation.
 
Finally, we calculated the HDO emission in the outflow in a $11''$
beam.  We started with a uniform [HDO]/[H$_2$] abundance of $3 \times
10^{-10}$, similar to the envelope value. This yields a spectrum which
matches the maximum intensity of the observed emission, but not the
wings (Fig. \ref{f15}). This may be due to the presence of multiple
outflows, similar to the H$_2$O case. The outer wings can be fitted
for an abundance of $3 \times 10^{-9}$ and results in a self-reversed
profile, but its peak intensity is too high by a factor of 2--4. This
is not surprising since we do not include an infalling envelope.  A
better fit may be obtained by a combined (multiple) outflow and infall
model as discussed above for H$_2$O.  We conclude that our shock model
can explain the wings of the observed HDO emission and that the HDO
abundance in the outflow may be somewhere in the range from 
$3\times 10^{-10}$ to $3\times 10^{-9}$.

\section{The [HDO]/[H$_2$O] Abundance Ratio Structure Toward 
I16293A\label{hdo_ratio}}

\subsection{Derived [HDO]/[H$_2$O] Profile From Envelope Excitation 
Analysis\label{hdo_h2o}}
 
It is clear that great caution should be taken when comparing the
H$_2$O abundance derived from the SWAS measurements and the HDO
abundance derived from the JCMT observations. Nevertheless we are able
to use the same models since we convolve the emerging emission with
the appropriate beam.  Because of the large beamsize, SWAS is not
sensitive to the H$_2$O emission from the warm inner envelope. To test
this we enhanced the abundance of H$_2$O for $T_{\rm kin} >$ 90 K to
the value of $10^{-4}$. This was not found to have any significant
impact on the emission in the SWAS beam. Clearly SWAS is mostly
sensitive to H$_2$O emission from extended colder gas in the outflow,
and to a lesser extent to gas in the outer envelope region (\S
\ref{outflow_model}).

The shocks associated with the outflows are able to release H$_2$O
from the grains and effectively convert oxygen into H$_2$O through
neutral-neutral reactions, but the velocities are so small that H$_2$O
formation at an $1\times 10^{-4}$ abundance level is
inhibited. Indeed, ISO LWS observations yield much lower H$_2$O
abundances. The ISO observations of H$_2$O are all from higher level
transitions $E/k > 100$ K, and thus trace the hot/warm disk/core
region of small extent. \citet{cec00} derive an [H$_2$O]/[H$_2$]
abundance in the `outer envelope' ($r\ge 150$ AU) of about $2.5 \times
10^{-7}$ which is similar to our envelope abundance ($2 \times
10^{-7}$), while for the inner region ($r\le 150$ AU) they derive an
enhanced abundance of $1.5 \times 10^{-6}$ which is attributed to
evaporation of water ice mantles of grains at $T_{\rm kin}\simeq 100$
K. The latter value is higher than our inferred outflow abundance of
about $10^{-8}$ (\S \ref{outflow_model}).

If we combine the H$_2$O abundances of \citet{cec00} with our derived
HDO abundances we find that the [HDO]/[H$_2$O] abundance ratio varies
between $2 \times 10^{-4}$ in the warm envelope, $r< 150$ AU,
$10^{-3}$ in the envelope with $r>150$ AU and $T_{\rm kin} > 14$ K,
and $8 \times 10^{-2}$ in the outer envelope where $T_{\rm kin} < 14$
K.  For a detailed comparison between the HDO and H$_2$O abundances
throughout the envelope we consider in the next section the
temperature effects of ion-molecule chemistry and the effects of
depletion of molecules and atoms which play a vital role in the
formation of HDO and H$_2$O.

\subsection{A Simple Chemical Model}
\label{Simple_chem_model}

In ion-molecule chemistry (e.g., \citealt{Pin89, Mil91}) HDO is
thought to be formed through the dissociative recombination
\begin{equation} 
{\rm H}_2{\rm DO}^+ + e^- \rightarrow {\rm HDO} +{\rm H},
\label{e4}
\end{equation}
with a statistical branching ratio of \case{2}{3} compared to the
corresponding recombination of H$_3$O$^+$. In steady state the
gas-phase density of H$_2$DO$^+$ can be written as
\begin{equation}
n({\rm H}_2{\rm DO}^+)=
{{(n({\rm O}) k_1 + n({\rm H}_2{\rm O}) k_2 ) n({\rm H}_2{\rm D}^+)
 + n({\rm H}_2{\rm O}) n({\rm DCO}^+) k_3} \over {n(e) k_4}},
\label{e5}
\end{equation}
where $k_i$ are the relevant reaction rate coefficients. We adopt
equal total rate coefficients for the reactions involving the
deuterated and hydrogenated species and assume statistical branching
ratios.  H$_3$O$^+$ is formed through similar reactions with
hydrogenated species.  We will neglect the formation of H$_3$O$^+$
from channels leaving the H$_2$DO$^+$ reaction sequence and vice
versa, and neglect the formation of H$_3$O$^+$ and H$_2$DO$^+$ from
reactions of H$_3^+$ or H$_2$D$^+$ with HCO$^+$, and DCO$^+$,
respectively.  The [HDO]/[H$_2$O] abundance can then be written
\begin{eqnarray} 
{x({\rm HDO}) \over {x({\rm H}_2{\rm O})}}& =
 & {2\over 3} {x({\rm H}_2{\rm DO}^+) \over {x({\rm H}_3{\rm O}^+)}}
\label{e6}\\ 
& \simeq& {1\over 18} {{x({\rm H}_2{\rm D}^+)} \over {x({\rm H}_3^+)}}.
\label{e7}
\end{eqnarray}

The heart of the fractionation of deuterated molecules is the
molecular-ion H$_2$D$^+$ which is formed through reaction
(\ref{e1}). In equilibrium the formation and destruction rates of
H$_2$D$^+$ are balanced which leads to \citep{Sta99}
\begin{equation}
{x({\rm H}_2{\rm D}^+) \over {x({\rm H}_3^+)}} = 
 {{ x({\rm HD})k_f + x({\rm D}) k_{\rm D}} \over
        {  x(e) k_e + \sum k_i x({\rm X}) + k_r}},
\label{e8}
\end{equation}
where $k_e$ is the dissociative recombination rate coefficient of
H$_2$D$^+$, $k_{\rm D} $ is the rate coefficient for the formation of
H$_2$D$^+$ via the reaction H$_3^+ + {\rm D}$, $x({\rm X})$ is the
fractional abundance of species X=CO, O, H$_2$O, etc., and $k_f$ and
$k_r$ are the forward and backward rate coefficients of reaction
(\ref{e1}), see \citet{Sta99} for references of the adopted rate
coefficients. The H$_3^+$ density depends on the effective cosmic ray
ionization rate $\zeta_{\rm eff}$ and can be written as \citep{Lep87}
\begin{equation}
n({\rm H}_3^+)={{\zeta_{\rm eff} \over {\sum k_i x({\rm X})}}}.
\label{e9}
\end{equation}

We consider only the destruction of H$_2$D$^+$ through reactions with
CO, assume $x({\rm HD})=10x({\rm D})=2.8 \times 10^{-5}$, and
$\zeta_{\rm eff}=5 \times 10^{-17}$ s$^{-1}$. Together with the
balance for H$_3^+$ formation and destruction (Eq.[\ref{e9}]) we
derive an expression for the H$_2$D$^+$ abundance with the CO
abundance as the only free parameter. This abundance structure was
subsequently used as input in the excitation calculations to compute
the spectrum of the ortho-H$_2$D$^+$ ground-state transition. A good
fit to the observed H$_2$D$^+$ spectrum is established for a two step
CO abundance profile with [CO]/[H$_2$]=$1.5 \times 10^{-4}$ for
$T_{\rm kin}>$ 22 K and $3\times 10^{-6}$ for $T_{\rm kin}< 22$ K
(Fig. \ref{f16}) applied to the infall envelope model.  This indicates
a CO-depletion by a factor of 50 at temperatures below 22~K, where CO
is frozen out. This value equals the depletion derived from our
C$^{17}$O 2--1 excitation analysis, but is a factor of 50 larger than
derived from our C$^{18}$O 2--1 and 3--2 observations which indicated
no depletion (\S \ref{envelope}). However, the spectra of the latter
transitions are biased toward the warm inner regions and the outflow.

Now we can use the [H$_2$D$^+]/[$H$_3^+$]-model fit, together with the
H$_2$O abundances from the excitation analysis and the expression of
the [HDO]/[H$_2$O] ratio (Eq. [\ref{e7}]) to determine an HDO model
abundance profile and to calculate a model spectrum analogous to
H$_2$D$^+$. The model spectra are shown in Fig. \ref{f16}.  When we
adjust [H$_2$O]/[H$_2$]=$1.25 \times 10^{-7}$ for $T_{\rm kin}>14$ K
and $2 \times 10^{-9}$ for $T_{\rm kin}< 14$ K we get a fit to the
observed emission peaks.  Note that these abundances agree within the
uncertainties with the derived H$_2$O excitation results.  The HDO
chemistry model spectrum (Fig. \ref{f16}) resembles the HDO excitation
infall spectrum (Fig. \ref{f8}) very well. The fact that the
absorption does not go down to zero is a result of the uncertainty in
the HDO collision rates for $T_{\rm kin}< 50$ K.  Note that the HDO
spectrum is calculated self-consistently from the theoretical
[HDO]/[H$_2$O] expression and the H$_2$O 2-value abundance profile
from the excitation analysis, i.e., it does not provide an
independently determined HDO abundance structure.

\subsubsection{Observational Results Versus Chemistry Models\label{vs}}

The upper-limits on the ortho-H$_3$O$^+$ abundances derived from the
observations by \citet{Phi92} in \S \ref{envelope} can be used to
determine an upper limit to the H$_2$O abundance. From the formation
and destruction balance of H$_2$O, we derive
[H$_2$O]/[H$_3$O$^+$]$\simeq 1000$--1500 where we used the
experimentally determined branching ratio of 0.5 from \citet{Jen00} to
form H$_2$O in the dissociative recombination of
H$_3$O$^+$. Furthermore we only considered the formation of H$_2$O via
the H$_3^+$ channel and assume that 10\% of all ions cause destructive
dissociation of H$_3$O$^+$, similar to \citet{Phi92}. The upper- and
lower-limits correspond to the $T_{\rm kin}\le 50$ K and $T_{\rm
kin}=80$ K ranges, respectively. The observations then yield
[H$_2$O]/[H$_2$]$< 5 \times 10^{-6}$ for both temperature ranges and
are consistent with the abundances derived by \citet{cec00} from ISO
observations.

The derived H$_2$O abundance in the outflow is $ 1.3\times 10^{-8}$
and agrees with the dynamical/chemical model of \citet{Ber99} for a
pre-shock chemistry which evolves until $t=10^{3}$ yr.  The same model
predicts significant deuterium fractionation in the pre-shock gas
after $10^6$ yr: [HDO]/[H$_2$O]$\sim 10^{-3}$, in agreement with our
derived value for the envelope region where $r>150$ AU and $T_{\rm
kin}>14$ K. For the warm inner-envelope region $r<150$ AU, we derive
[HDO]/[H$_2$O]=$2 \times 10^{-4}$. Our gas-phase chemistry model
predicts values about two orders of magnitude lower for this
region. We attribute the observed enhancement to the shock induced
release of solid water (HDO, H$_2$O) from the grains.

\subsubsection{[HDO]/[H$_2$O] in the Cycle of Star- and Planet 
Formation\label{cycle}}

Our derived [HDO]/[H$_2$O] ratio in the cold ($T_{\rm kin}<14$ K)
envelope regions is $0.08$. This ratio is however an upper limit due
to the fact that the HDO collision rates are only available for
$T_{\rm kin}\ge 50$ K. For lower temperatures we used the 50~K values,
but the rates are expected to decrease for lower temperatures, so that
the true HDO abundance may be lower than we derived in \S
\ref{CO_outflows}.  Since oxygen starts to freeze out at $T_{\rm
kin}=14$ K we may also expect a drop in HDO which is similar to H$_2$O
(\S \ref{Simple_chem_model}. Thus, the [HDO]/[H$_2$O] ratio may be in
the range $10^{-3}$--$8 \times 10^{-2}$ at $T_{\rm kin}< 14$ K.  The
[HDO]/[H$_2$O] chemistry model (Eq. [\ref{e7}]) reaches a maximum
value of $\sim 0.06$ when [H$_2$D$^+$]/[H$_3^+$]$\rightarrow 1$.
Thus, the observed [HDO]/[H$_2$O] ratios can be understood from pure
gas-phase deuterium chemistry at low temperatures.

The [HDO]/[H$_2$O] abundance ratio in the warm inner-envelope $(2
\times 10^{-4})$ of I16293A is close to the value that has been
measured in the solar system: $3.0 \times 10^{-4}$ in the comets
Hale-Bopp, Halley, \& Hyakutake \citep{Ebe95, Boc98, Mei98}, and $ 1.5
\times 10^{-4}$ in the Earth's oceans \citep{Lec98}.  These values are
all so close as to suggest that the [HDO]/[H$_2$O] abundance ratio may
be a constant in the cycle of star- and planet formation.  The
fundamental question is then: where and when is the [HDO]/[H$_2$O]
ratio determined?  There are currently two favoured scenarios (e.g.,
\citealt{Aik99, Aik01}): 1) deuteration from ion-molecule chemistry in
the protoplanetary disk; 2) deuteration determined by gas phase and/or
solid state chemistry in the dense interstellar medium, namely in the
parent molecular cloud before star formation begins.  \citet{Aik02}
modeled the [HDO]/[H$_2$O] ratio in accretion disks models and found
values of the order of $10^{-3}$ for $R>100$ AU in the cold midplane
where $T_{\rm kin}\simeq 15$ K.  This may indicate a large depletion
of H$_2$O while HDO is still in the gas phase.  Unfortunately the JCMT
and SWAS obervations cannot be used to determine the (deuterated)
water abundances in the circumstellar disk.  For the moment we can
only make a comparison for the HDO abundance in the warm envelope. Our
derived [HDO]/[H$_2$] abundance ratio of $3\times 10^{-10}$ is
comparable within a factor of a few to the protoplanetary disk model
of \citet{Aik02} with mass accretion rate $10^{-9}~ M_\sun$ yr$^{-1}$,
and supports the hypothesis that the HDO abundance in the disk and
warm envelope are equal.

\section{The nature of I16293A,\,B and I16293E}
\label{Nature_of}

\subsection{I16293A\label{i16293a}}

I16293A is a Class~0 source, i.e., is believed to be a protostar
according to the classification proposed by \citet{And93}.  It is
invisible in the near- and the mid-infrared, it is associated with
free-free emission so it has already formed a stellar-like core, it
drives a powerful molecular outflow, and it is surrounded by an
envelope of dust and gas.  Even though the temperature that we derive
for the dust disk surrounding I16293A is higher than the canonical
value of 30~K proposed by \citet{And93}, the ratio of submillimeter to
bolometric luminosity is typical for Class~0 sources, namely $> 5
\times 10^{-3}$. For I16293A we derive ${\rm L_{submm}/L_{bol}}\geq 2
\times 10^{-2}$. Note that this is really a lower limit, since our
estimate for the bolometric luminosity also includes the luminosity of
I16293B, which is likely to be a few \Lsun, while its contribution to
the submillimeter luminosity is negligible.  \citet{Bon96}, in their
study of outflow properties of a sample of Class~I and Class~0
sources, find that Class~0 sources in general have a much higher
outflow efficiency (defined as the ratio of momentum flux to radiative
force) than Class~I sources, with an average ratio for Class~0 sources
of about 1000. For the I16293A outflow we find $F_{\rm CO}/F_{\rm
rad}\sim 730$, which is close to the average value found by
\citet{Bon96}.  They also included I16293A in their sample, but the
outflow momentum they derive is significantly larger because they also
include the E--W outflow, which is driven by I16293B.

\subsection{I16293B\label{i16293b}}

Although it is most likely that I16293A and I16293B originate from the
same collapsing core, I16293B remains much more of an enigma. Our
observations support the suggestion of \citet{Walk93} that I16293B
drives the large E--W outflow. Our observations also show that the
submillimeter dust emission peaks on I16293A and more or less follows
the C$^{18}$O emission mapped by \citet{Mun90}. Single dish
observations have insufficient resolution to show any contribution
from I16293B at all, even though we know from VLA and 3~mm aperture
synthesis studies (e.g., \citealt{Mun92}) that the spectral index of
I16293B is much steeper than that of I16293A and indicative of dust
emission.

What is I16293B? Based on the scarce evidence that we have, I16293B
looks like a young T~Tauri star, which has driven an extremely
powerful large outflow which extends $\sim 0.45$ pc on the
blue-shifted side. Such outflows from low-luminosity young stars are
not uncommon; they are seen both as large scale optical jets
\citep{Bal97} and large scale CO outflows \citep{Pad97}. I16293B is a
low-luminosity star, $< 5$ \Lsun, associated with compact dust
emission; the true luminosity may be found by means of high resolution
millimeter-wave observations.

\subsection{I16293E\label{i16293e}}

We derive a bolometric luminosity $L_{\rm bol}=2.3$ \Lsun\ for the
prestellar core I16293E, and find that its FIR continuum spectrum
resembles that of the Class~0 source I16293A (Fig. \ref{f4}).
\citet{Cas01} quote a dust temperature $T_{\rm d}=24$ K for this
source, higher than our value of $T_{\rm d}=16$ K since we correct for
the overlapping envelope of I16293A at the position of I16293E. This
correction is important since both cores are connected with a
bridge-like structure (Figs. \ref{f1}, \ref{f5}).

The temperature and density structure of I16293E resembles the
conditions in the outer envelope of I16293A. It is very interesting
that the HDO abundance of the warm envelope of the YSO, and the core
of I16293E are similar. This suggests that some of the gas species in
the warm envelope may still have the abundance that they had in the
prestellar stage, the difference being that the gas in the warm
envelope is excited and thus easier to detect. A comparison between
I16293E and I16293A could therefore be very important to understand
the formation of a first hydrostatic core, a so-called Class~$-$I
protostar \citep{Bos95}. Such protostars lie between the prestellar
clouds and the Class~0 protostars, and occur when a collapsing cloud
first becomes optically thick, heats up and reaches a quasi continuum
stage. This stage is short lived, about $2\times 10^4$ yrs, after
which a second collapse takes place and the final protostar is formed.
Class~$-$I protostars are characterized by a central core of about
200~K, embedded in a cold core of about 10~K, although the models of
\citeauthor{Bos95} mostly reached a central source with a much lower
temperature of 20~K.

Is a first protostar present in I16293E? Most of the observed
molecular line profiles from I16293E show two peaks with a stronger
redshifted peak, characteristic of expansion motions.  In \S \ref{ew}
we argue that this may be due to projection effects of pieces of the
E--W ouflow from I16293B which streams around I16293E.  We also do not
find an outflow associated with I16293E in our CO maps (\S \ref{ew}),
so it should be younger than a Class~0 object.  A central Class~$-$I
protostar with $T_{\rm kin}=20$--200 K may be present but very hard to
detect in deep mid- or far infrared continuum imaging since the
spectral signature will be only a weak shoulder at the Wien side of
the energy spectrum of the core.  Such a first protostar will be very
compact and have little mass, $M_1\le 0.04~M_\sun$, which can easily
be hidden in the dense cold I16293E core.  Detection with SCUBA seems
impossible because of the beam size and because the outer envelope of
such a cold core will be warmer due to the interstellar radiation
field and will dominate the continuum emission. Our HDO emission,
which we think comes from the heart of the core, has a narrow
linewidth ($\Delta V= 0.6$ km~s$^{-1}$) and our models yield a kinetic
temperature of about 16~K. This indicates that a first protostar must
still be rather cold at the low end of the temperature interval of
\citet{Bos95}.  Sensitive high resolution submillimeter interferometry
of the continuum with, e.g., the Atacama Large Millimeter Array may
reveal a cold accretion disk.

\subsection{The Extreme Deuteration of LDN~1689N\label{deuter}}

We think that LDN~1689N is a remarkable low-mass star-forming region,
but not because it `harbors a young protostar, two molecular outflows
and a cold nearby molecular core' \citep{Lis02}. Multiple outflows are
common around YSOs and have been reported, e.g., in L1448
\citep{bar98, WC00}, HH~24 \citep{Eis97}, and NGC~1333
\citep{kne00}. Deep studies of embedded YSOs in low-mass cores
revealed the presence of apparently starless cold companion cores with
masses of about a few $M_\sun$ \citep{Mot98, Hog00a, Shi00}.  Strong
deuteration is expected in cold dense prestellar cores which are on
the verge of forming a first hydrostatic core because of the extreme
depletion of molecules like CO which may result in an abundace ratio
[H$_2$D$^+$]/[H$_3^+$]$> 1$ and a boost of the [D]/[H] ratio in
molecules (\S \ref{s:intro}). Indeed, several prestellar clouds have
now been found to have a large depletion of CO (e.g., L1544,
\citealt{cas02}; L68, \citealt{Ber02}) and a large H$_2$D$^+$
abundance \citep{cas03}.  But, what really is remarkable in LDN~1689N
is that it can be considered as a `deuteration candle' with the
strongest deuteration known so-far in low-mass YSOs.  What causes the
extreme deuteration in the envelope of I16923A and the core of
I16293E? The answer may lie in the fact that I16293AB is a binary
system where the YSOs and their outflows differ in age.  I16293B is
probably the oldest of the two YSOs and the source of the fossilized
E--W outflow (\S \ref{ew}). This outflow is responsible for the
dynamical interaction of a part of the eastern blue lobe with a dense
dark core, the precursor of I16293E. This interaction caused a shock
with velocity $V_{\rm s}=8$--10 km~s$^{-1}$ \citep{Lis02} as measured
at the deuterium peak D north of I16293E and possibly higher at the
first place of impact west of the core. This could have released
deuterated species which were condensed on the dust grains, and
compressed the gas, which then cools efficiently to low temperature
\citep{Lis02}. Since the E--W outflow causes a less energetic shock
than the NE--SW outflow \citep{Lis02, Gar02} and SiO emission is found
only associated with the NE--SW outflow (e.g., \citealt{Gar02}), we
expect only deuterated species to desorb since these were frozen out
relatively late and thus are located near the outer surface of the
grain mantles. These slow shocks of the outflow leave the grains
mostly intact, with a post shock composition like that of a grain in a
$T_{\rm kin}=20$ K gas. The release of deuterated species, combined
with the gas-phase deuterium chemistry at low temperatures, boosts the
deuterium enrichment to extremely high values just after the passage
of a shock; these deuterated species freeze out again in about $10^4$
yrs.

Compression from this shock resulted in an elongated N--S structure,
visible in the 450 $\mu$m image (Fig. \ref{f5}). At the north side
this filamentary structure bends to the northeast where the gentle
flow passes around the core; here the deuterium peak is located. This
shock would cause a blue-shift of the systemic velocity in I16293E
with respect to the parental LDN~1689N cloud. Indeed, the HDO spectrum
toward I16293E is blue-shifted by about 0.3 km s$^{-1}$ with respect
to the systemic velocity of the LDN~1689N envelope toward I16293A
(Table \ref{t1}). \citet{Lis02} also found such a blue-shift of
deuterated molecules toward the deuterium peak D which is located
about 10 arcsec northwest of the continuum peak of I16293E. We believe
that at this position the shock has the greatest chemical impact.

This strong deuterated core may have been shocked again at a later
time by the NE--SW outflow from I16293A causing a velocity red shift
of 0.3 km~s$^{-1}$.  However, only the non-deuterated species, which
predominently reside in the outer core region, are red-shifted with
respect to their deuterated forms found mainly in the heart of the
core. This is exactly the velocity shift which \citet{Hir01} would
expect from an interaction of the northeast red lobe. However, they
did not detect this, since they measured only the H$^{13}$CO$^+$ 1--0
line at the position of I16293E which has a velocity equal to the
cloud systemic velocity. Under our hypothesis this has to be measured
with respect to the reference frame of the first shock.
 
This second shock boosted the gas-phase deuterium chemistry to its
maximum and the temperature is further lowered due to further
compression.  We think that it is this firecracker effect which leads
to the extreme deuteration of LDN~1689N in general, and I16293E in
particular. Multiple shocks may be common in low-mass YSOs;
\citet{Lar02} suggest that Serpens SMM1 is influenced by multiple
low-velocity shocks driven by the outflow.

\section{Discussion and Conclusions\label{conclusion}}

We have analyzed our submillimeter dust and continuum data in terms of
an excitatiom analysis using a model for the temperature and density
structure of the disk, envelope, and outflow of I16293A, B, and E. The
spectral lines of H$_2$O, HDO, and H$_2$D$^+$ were subsequently
compared with predictions from a simple chemical network.

Figure \ref{f17} shows a schematic overview of the IRAS 16293-2422 
region showing its distinct physical regions (disk, warm dense core, 
warm envelope, cold envelope, outflow regions, etc.). For each region 
we indicate the derived physical parameters, and their molecular 
abundances. For the submillimeter disk of I16293A we derive a dust 
temperture $T_{\rm d}=40$ K, a source size of about 800 AU, a total 
mass of 1.8\Msun\ and a luminosity of 16.5 \Lsun\ which corresponds to 
a gas density $n({\rm H}_2)\ge 10^9$ cm$^{-3}$.  The envelope of I16293A 
can be well described by a collapsing envelope model with parameters age
$t=(0.6$--$2.5) \times 10^4$ yr and sound speed $a=0.7$ km s$^{-1}$.
In fact the center of the expansion wave, $r_{\rm CEW}= a t$, is
890--3700~AU from the centre which corresponds to 6--23$''$ at the
distance of 160 pc, comparable to the beam size of the SCUBA continuum
and submillimeter line emission of the species we used in the
analysis.  The outer radius of the envelope is about 7300 AU and the
temparature ranges from 115 K in the inner region to 12 K in the outer
region. The envelope mass is $M_{\rm env}({\rm H}_2)=6.1$ \Msun. We
find that for a YSO such as I16293A the line emission is beter suited
than the continuum emission to constrain the infall parameters. The
excitation analysis of the envelope yields a constant HDO abundance of
$3\times 10^{-10}$ throuhout the envelope, and an ortho-H$_2$O
abundance of $2\times 10^{-7}$ for the regions with $T_{\rm kin}>14$
K, while the abundance is lowered by a factor of 150 in the colder
regions. From the C$^{17}$O analysis we find that CO is depleted by a
facot of 50 for the envelope regions where $T_{\rm kin}<20$ K.  The
line wings of the CO, HDO, and H$_2$O emission can qualitatively be
explained by a single outflow model with a turbulent velocity of 2.3
km~s$^{-1}$ and where the maximum outflow velocity is about 6
km~s$^{-1}$.  The excitation analysis of the outflow model indicates
that the ortho-H$_2$O abundance in the outflow is about $1 \times
10^{-8}$ and that the HDO abundance in the outflow is in the range of
$3\times 10^{-9}$ to $3 \times 10^{-10}$.  A more quantitative
comparison requires a multiple outflow model.  We find that only two
outflows are present, a NE--SW outflow with a dynamical time scale of
about $(3.0$--$3.5) \times 10^3$ yr which is powered by I16293A a
Class~0 source, and an E--W outflow which is a fossil flow with a
dynamical time scale of $(6.4$--$7.6) \times 10^3$ yr driven by
I16293B which looks like a young low-luminosity T~Tauri star.  The
prestellar object I16293E is well described with an isothermal core of
$T_{\rm kin}=16$ K, a core radius of 8000 AU and a power law density
structure of the form $n(r)=n_0 (r/1000~{\rm AU})^{-p}$ where
$n_0=n(R=1000$ AU$)=1.6 \times 10^6$ cm$^{-3}$. The total mass mass of
this core is $M_{\rm core}({\rm H}_2)=4.35$ \Msun.  Our excitation
analysis of the core indicate a depletion of CO and HCO$^+$ by a
factor of ten relative to the `standard abundances' and that the
DCO$^+$ abundance is larger than that of H$^{13}$CO$^+$. Our tentative
detection of HDO indicates the presence of a cold condensation in the
heart of this core where the HDO abundance is about $2\times
10^{-10}$.  Our detailed CO mapping shows that I16293E has no outflow,
so its evolutionary stage is younger than Class~0. We argue that this
core may hide a first hydrostatic core, a so called Class~$-$I object
which is deeply embedded in a largely unaffected cold core.

The reason for the extreme deuteration in the L16293N cloud and in 
particular in I16293A, and I16293E may lie in the fact that I16293AB 
is a binary system where the YSOs and the outflows differ in age 
causing a firecracker efffect of the deuteration. In this scenario 
the older
E--W ouflow from I16293B is responsible for the formation of I16293E
out of a dense dark core through a slow shock where only deuterated
species were desorbed from the grains which, in combination with the
low temperatures, amplifies the deuterium enrichment in the gas phase
before a possible re-absorption on grains. This strong deuterated core
may have been shocked again an a later time by the younger NW--SW
outflow from I16293A further boosting the gas-phase deuterium
chemistry to extreme values.

It is remarkable that the abundances of many molecules in the
prestellar core I16293E and in the warm part of the envelope of
I16293A are comparable. In particular, the HDO abundances are similar
which indicates that the warm gas in I16293A may have gone through the
same cold pre-collapse phase that I16293E is currently in.  A simple
chemical gas-phase modeling of the deuterium chemistry requires that
CO is depleted by a facor of 50 for $T_{\rm kin}<22$ K in order to
reproduce the observed H$_2$D$^+$ emission from our chemical network
applied to the envelope $(T,\,n)$ structure, and that an H$_2$O
abundance of $3 \times 10^{-7}$ for $T_{\rm kin}>14$ K and $4\times
10^{-9}$ for $T_{\rm kin}<14$ K is required to reproduce the observed
spectrum from the chemical model. The [HDO]/[H$_2$O] abundance in the
warm inner-envelope of I16293A ($2\times 10^{-4}$) is close to the
value that has been measured in the solar system and supports
this. Our chemical modeling indicates that the [HDO]/[H$_2$O] ratio is
determined in the cold gas phase prior to star formation, and that
this ratio is conserved in the early stages of low-mass star through
freeze out on dust grains formation and that the molecules are
released unprocessed to the gas phase after the formation of a
low-mass YSO. If this warm deuterated gas is subsequently locked-up in
a proto-planetary disk before the high temperature gas chemistry
causes a de-deuteration, the [D]/[H] ratio of HDO in low-mass
pre-solar systems would remain conserved.  To investigate this
scenario in detail, sensitive follow-up observations of the HDO and
H$_2$O ground-state transitions in a large sample of low-mass YSOs are
needed at high spatial and spectral resolution (i.e., interferometry
from high altitudes [HDO] and from space [H$_2$O]) and allow a study
of the [HDO]/[H$_2$O] abundance evolution from prestellar cores to
planets.  The H$_2^{18}$O ground-state transition may be more suitable
than the ortho-H$_2$O transition since the ground-state of the latter
molecule is at $\Delta E/k=34.3$ K above ground, while that of
H$_2^{18}$O is at 0~K, and the main isotopic form of water is expected
to be mostly optically thick in its ortho and para ground-state lines.

\acknowledgements The James Clerk Maxwell Telescope is operated by the
Joint Astronomy Centre, on behalf of the Particle Physics and
Astronomy Research Council of the United Kingdom, the Netherlands
Organisation for Scientific Research, and the National Research
Council of Canada.  It is a pleasure to thank the JCMT staff and the
MPIfR Division for Submillimeter Technology for their outstanding
support.  The Canadian Astronomy Data Center is operated by the
Dominian Astrophysical Observatory for the National Research Council
of Canada's Herzberg Institute of Astrophysics.

%%%%%%%%%%%%%%%%%%%%%%%%%%%%%%%%%%%%%%%%%%%%%%%%%%%%%%%%%%%%%%%%%%%%%%

%%%%%%%%%%%%%%%%%%%%%%%%%%%%%%%%%%%%%%%%%%%%%%%%%%%%%%%%%%%%%%%%%%%%%%

\clearpage

%%% Table 1  ! in AASTeX format !

\begin{deluxetable}{lrrr}
  \tablecaption{Submillimeter Photometry of I16293E and Total Flux
    Densities of I16293A,\,B Derived From Maps\label{t1}}
  \tablehead{
    & \colhead{I16293E} & \colhead{I16293AB} \\
    & \colhead{Flux Density} & \colhead{Integrated Flux} &
    \colhead{Envelope} \\
    \colhead{Filter/HPBW} & \colhead{(Jy~beam$^{-1}$)} & \colhead{(Jy~beam$^{-1}$)} &
    \colhead{(Jy)}
  }
  \startdata
  2.0\,mm/27\asec\ (2)   & $0.20 \pm 0.03$ & \\
  1.3\,mm/19.5\asec\ (1) & $0.60 \pm 0.05$ & $7.0 \pm 1.4$\tablenotemark{a} & \\
  1.1\,mm/18.5\asec\ (3) & $0.70 \pm 0.07$ & $8.6 \pm 0.9$ & \\ 
  850\mic/14.5\asec     & \nodata & $21.9 \pm 2.2$ & 16.0 \\
  850\mic/14.0\asec     & $1.40 \pm  0.02$\tablenotemark{b} & $20.4 \pm 0.03$\tablenotemark{c} \\
  800\mic/16.5\asec\ (2) & $2.06 \pm 0.11$ & $24.5 \pm 2.5$ & \\
  750\mic/13.3\asec     & \nodata & $26.6 \pm 10.0$ & \\
  450\mic/18\asec\ (2)   & $11.8 \pm 2.2$ & $126 \pm 15$ & 102.7 \\
  450\mic/8.0\asec      & $4.35 \pm 0.09$\tablenotemark{b} & $126 \pm 0.2$\tablenotemark{c} \\
  350\mic/19\asec\ (1)   & $19.4 \pm 2.3$ & $175 \pm 35$ & \\
  \enddata
  \tablecomments{Numbers in parentheses denote the number of
    independent observations obtained for each filter/aperture
    combination.}
  \tablenotetext{a}{ from \citet{Mez92}.}
  \tablenotetext{b}{Peak flux density in cleaned SCUBA maps, see text.}
  \tablenotetext{c}{Integrated flux from Gaussian fit to SCUBA
  map. Errors are 1$\sigma$ and do not account for systematic
  calibration uncertainties.}
\end{deluxetable}

%%% Table 2

\begin{deluxetable}{lcrrrr}
\tabletypesize{\footnotesize}
  \tablecaption{Molecular Line Observations of I16293AB and
  I16293E\label{t2}}
  \tablehead{
    & & \colhead{$\int T dV$} & \colhead{$T$} &
    \colhead{$\Delta V$} & \colhead{$V_{\rm LSR}$} \\
    \colhead{Molecule} & \colhead{Transition} &
    \colhead{(K~km~s$^{-1}$)} & \colhead{(K)} &
    \colhead{(km~s$^{-1}$)} & \colhead{(km~s$^{-1}$)}
  }
  \startdata
  \tableline
  \multicolumn{6}{c}{I16293A\,B, in units of $T_{\rm MB}$}\\
  \tableline
  C$^{18}$O      & 3--2 & 33.3 & & & 4.0 \\
                 & 2--1 & 18.4 & 6.50 & 2.50 & 3.76 \\
  C$^{17}$O      & 2--1 & 9.3 & 3.24 & 2.86 & 3.99 \\
  $^{12}$CO      & 7--6 & 473 & & & 3.9 \\
                 & 6--5 & 408 & & & 3.9 \\
  HCO$^+$        & 4--3 & 81 & & & 4.1 \\
                 & 3--2 & 70 & & & 4.0 \\
  H$^{13}$CO$^+$ & 4--3 & 7.2 & & & 4.7 \\
  H$^{13}$CO$^+$ & 3--2\tablenotemark{a} & 11.8 & 4.87 & 2.27 & 4.23 \\
                 & 3--2\tablenotemark{b} & $-2.3$ & $-2.76$ & 0.78 & 4.24 \\         
  DCO$^+$        & 5--4 & 3.1 & & & 4.0 \\
                 & 3--2 & 5.18 & 2.56 & 2.02 & 4.48 \\
  H$_2$D$^+$     & $1_{10}$--$1_{11}$ & 0.8 & 1.1 & 0.6 & 3.9 \\
  N$_2$H$^+$     & 4--3 & 4.4 & 2.6 & 1.7 & 3.7 \\
  HDO            & $1_{01}$--$0_{00}$\tablenotemark{a} & 6.0 & 1.0 & 5.9 & 4.3 \\
                 & $1_{01}$--$0_{00}$\tablenotemark{b} & $-1.4$ & $-2.1$ & 0.60 & 4.2 \\
  H$_2$O         & $1_{10}$--$1_{01}$ & 2.9 & & & 3.9 \\
  \tableline
  \multicolumn{6}{c}{I16293E, in units of $T_R^*$}\\
  \tableline
  C$^{18}$O      & 2--1 & 5.40 & 3.60 & 1.41 & 3.80 \\
  C$^{17}$O      & 2--1 & 1.70 & 1.26 & 1.27 & 4.03 \\
  H$^{13}$CO$^+$ & 3--2 & 1.64 & 1.25 & 1.23 & 4.08 \\
  DCO$^+$        & 3--2 & 3.74 & 3.14 & 1.12 & 3.76 \\
  HDO            & $1_{10}$--$1_{11}$ & 0.1 & 0.15 & 0.6 & 3.5 \\
  \enddata
  \tablecomments{For those spectra where we only list the integrated
    intensity, $V_{\rm LSR}$ is the velocity of the self-absorption.}
  \tablenotetext{a}{Emission component.}
  \tablenotetext{b}{Absorption component.}
\end{deluxetable}

%%% Table 3

\begin{deluxetable}{lrrrrr}
  \tablecaption{Turbulent Line Width $b$ and Inferred Abundances for
  I16293AB and I16293E\label{t3}}
  \tablehead{
    & & \colhead{I16293A,\,B} & \colhead{I16293A,\,B} &
    \colhead{I16293A,\,B} & \\ 
    & \colhead{I16293E} & \colhead{$b$} & \colhead{Envelope} &
    \colhead{Envelope} & \colhead{Outflow}\\
    \colhead{Species} & \colhead{Abundance} & \colhead{(km~s$^{-1}$)}
    & \colhead{Abundance} & \colhead{Abundance\tablenotemark{a}} &
    \colhead{Abundance}
  }
\startdata
C$^{18}$O      & $3(-8)$ & 0.83 & $2(-7)$ & $6.2(-8)$ & \\
C$^{17}$O      & $9(-9)$ & 0.76 & $4(-8)$ & $1.6(-8)$ & \\
HCO$^+$        & $1(-10)$ & 0.69 & $<1(-9)$ & $1.4(-9)$ & \\
H$^{13}$CO$^+$ & $2(-11)$ & 0.50 & $2(-11)$ & $2.4(-11)$ & \\
DCO$^+$        & $5(-11)$ & 0.67 & $2(-11)$ & $1.3(-11)$ & \\
H$_2$O         & & & $3(-7)$\tablenotemark{b} & & $1.3(-8)$ \\
               & & & $4(-9)$\tablenotemark{c} & & \\
HDO            & $2(-10)$ & 0.20 & $3(-10)$ & & \\
H$_2$D$^+$     & & & $1(-12)$\tablenotemark{d} & & \\
               & & & $2(-9)$\tablenotemark{e} & & \\
N$_2$H$^+$     & & & $3(-11)$ & & \\
H$_3$O$^+$     & & &  $<5(-9)$ & & \\
\enddata
\tablecomments{$c(-d)$ denotes $c \times 10^{-d}$.}
\tablenotetext{a}{Abundance derived by \citet{Scho02} in a similar analysis.}
\tablenotetext{b}{Where $T_{\rm kin}>14$ K.} 
\tablenotetext{c}{Where $T_{\rm kin}<14$ K.} 
\tablenotetext{d}{Where $T_{\rm kin}>20$ K.} 
\tablenotetext{e}{Where $T_{\rm kin}<20$ K.} 
\end{deluxetable}

%%% Table 4

\begin{deluxetable}{lrrrr}
  \tablecaption{Physical Parameters of the NE--SW Outflow\label{t4}}
  \tablehead{
    & \multicolumn{2}{c}{Uncorrected\tablenotemark{a}} &
    \multicolumn{2}{c}{Corrected\tablenotemark{b}}\\
    \colhead{Parameter} & \colhead{SW Blue} & \colhead{NE Red} &
    \colhead{SW Blue} & \colhead{NE Red}
  }
  \startdata
  Dynamical time scale (yr)  &   $6.4 \times 10^3$     & $7.6 \times 10^3$ & $3.0 \times 10^3$     & $3.5 \times 10^3$ \\
  Mass (M$_\sun$)            & 0.10  &  0.05 & 0.33  & 0.18\\
  Momentum (M$_\sun$ km s$^{-1}$) & 0.60  & 0.22 & 2.27  & 1.25\\
  Energy (M$_\sun$ km$^2$ s$^{-2}$) & 3.44  & 0.78 & 23.4  & 8.64 \\
  Force (10$^{-5}$ M$_\sun$ km s$^{-1}$ yr$^{-1}$) & 1.91 & 0.74 & 6.43 & 4.90\\
  \enddata
  \tablenotetext{a}{Uncorrected for opacity and inclination.}
  \tablenotetext{b}{Corrected for opacity and inclination.}
\end{deluxetable}

%%%%%%%%%%%%%%%%%%%%%%%%%%%%%%%%%%%%%%%%%%%%%%%%%%%%%%%%%%%%%%%%%%%%%%

\clearpage

\begin{figure} % Figure 13
  \epsscale{1}
%\plotone{f13.eps}
  \caption{\small Velocity position contour plots of the two outflows driven
    by IRAS 16293--2422 A and B. The top figure is created by rotating
    the map with the position angle of the outflow $-35$\degree\ and
    integrating over a 20\arcsec\ wide strip approximately centered on
    A, i.e. this velocity position diagram goes along the symmetry
    axis of the NE--SW outflow, but also includes some contribution
    from the E--W outflow driven by B. The bottom plot is derived by
    rotating the CO map by 96\degree\ and integrating over a
    20\arcsec\ wide strip approximately centered on B.\label{f13}}
\end{figure}

\begin{figure} % Figure 14
  \epsscale{0.9}
\plotone{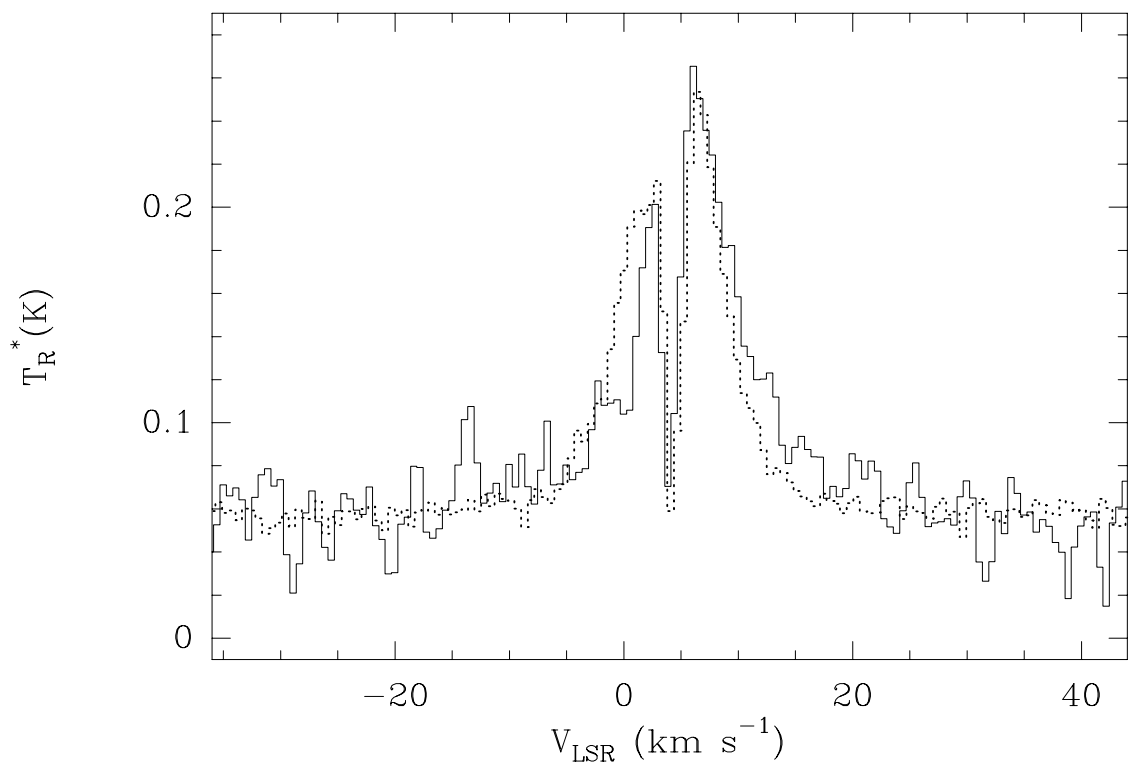}
  \caption{H$_2$O spectrum from SWAS ({\sl thick line}) superposed on
    the CO $J$=7--6 emission measured with the JCMT ({\sl dotted
    line}). The CO emision has smoothed to the same spectral
    resolution as SWAS and scaled down to match the peak of the H$_2$O
    emission.\label{f14}}
\end{figure}

\begin{figure} % Figure 15
  \epsscale{0.8}
\plotone{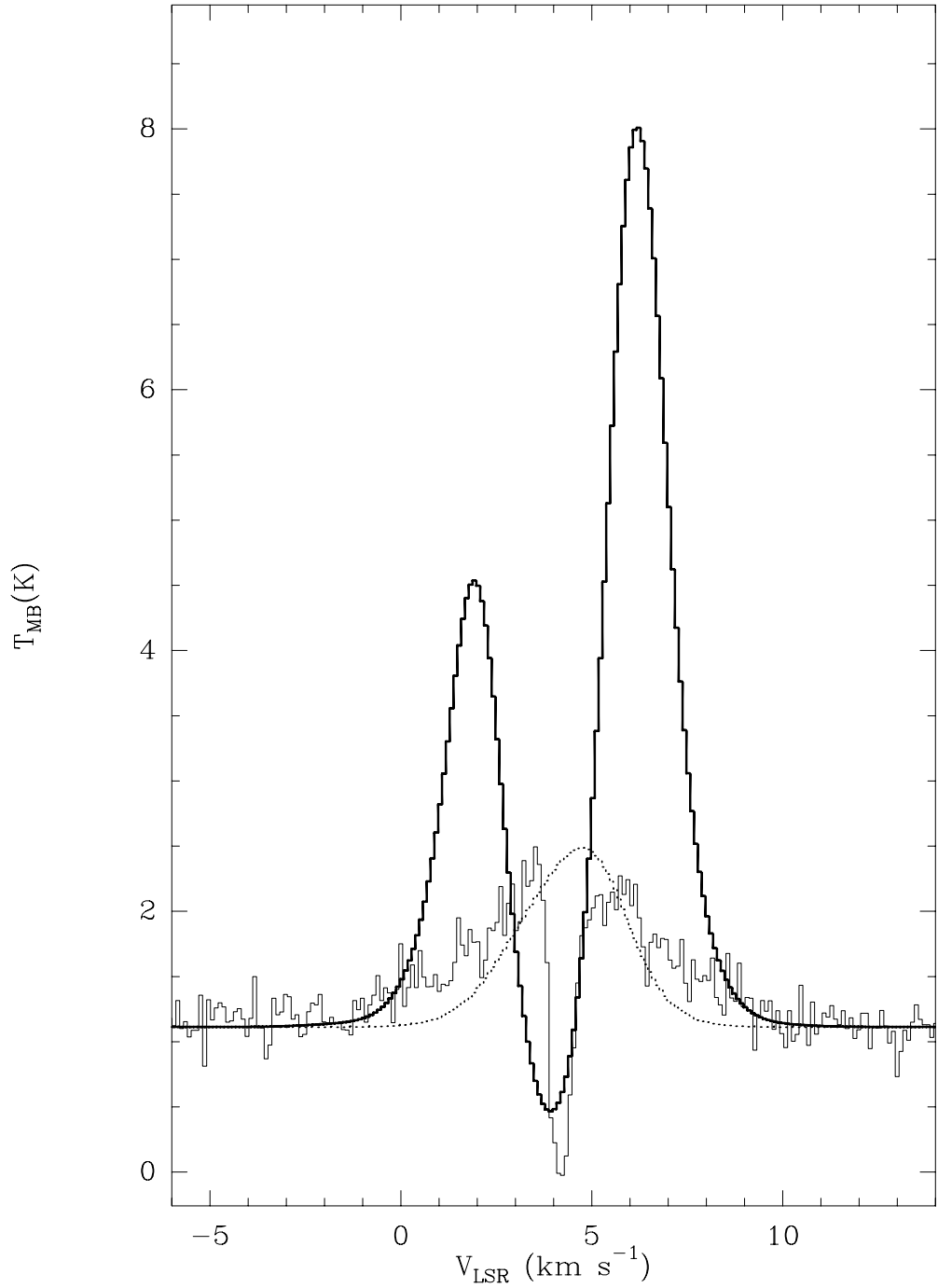}
  \caption{Spectrum of the observed HDO ground-state transition
    overlaid with outflow model spectra for [HDO]/[H$_2$] abundances of
    $3\times 10^{-10}$ ({\sl dotted}) which matches the level of the
    maximum emission, and $3\times 10^{-9}$ ({\sl thick}) which fits
    the outer wings.\label{f15}}
\end{figure}

\begin{figure} % Figure 16
  \epsscale{1.0}
\plotone{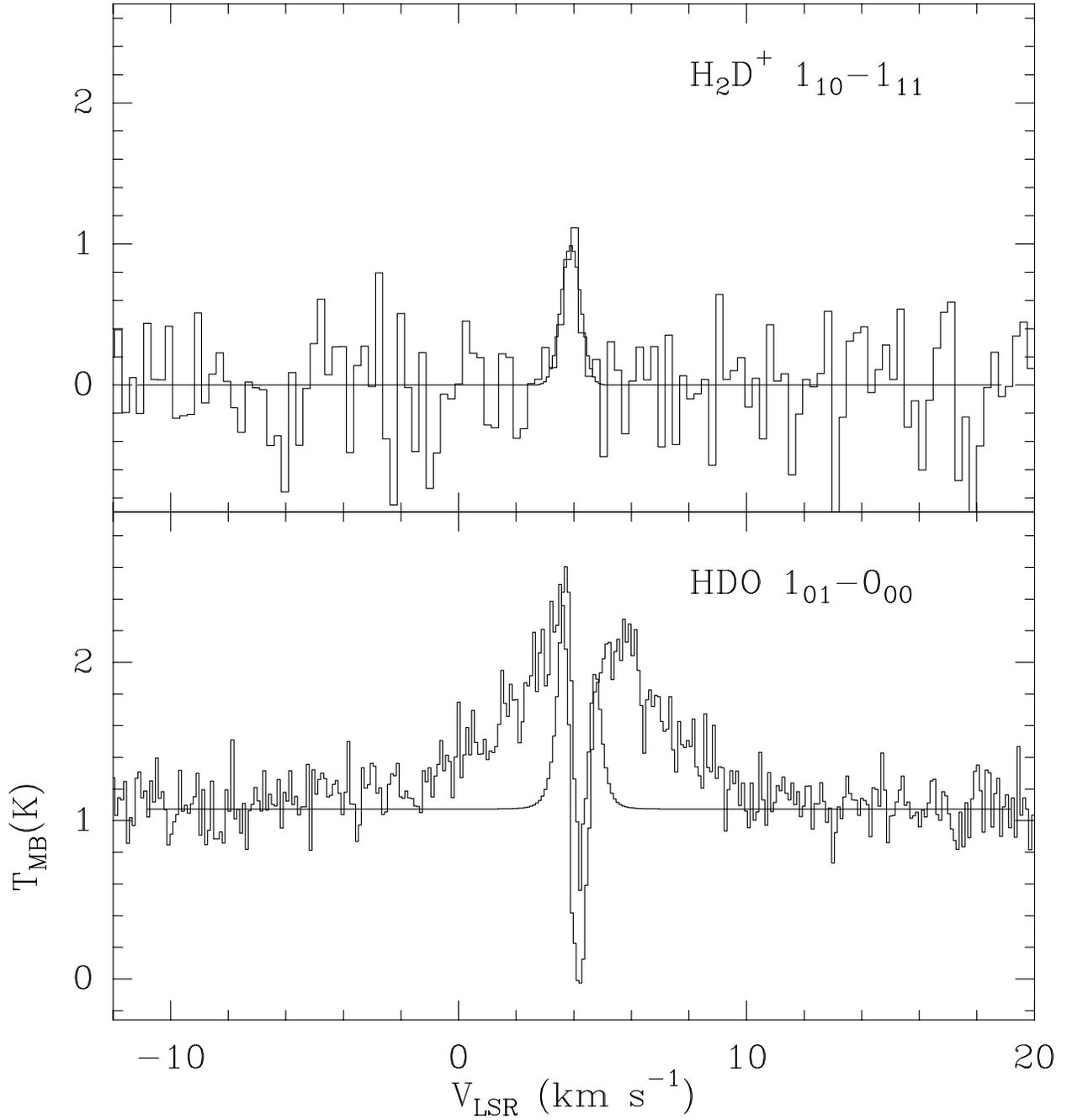}
  \caption{Model spectra obtained from the chemical calcuations for an
    infalling envelope, overlaid on the observed H$_2$D$^+$
    $1_{10}$--$1_{11}$ ({\sl top}) and HDO $1_{01}$--$0_{00}$ ({\sl
    bottom}) observations.\label{f16}}
\end{figure}

\begin{figure} % Figure 17
  \epsscale{1.0}
\plotone{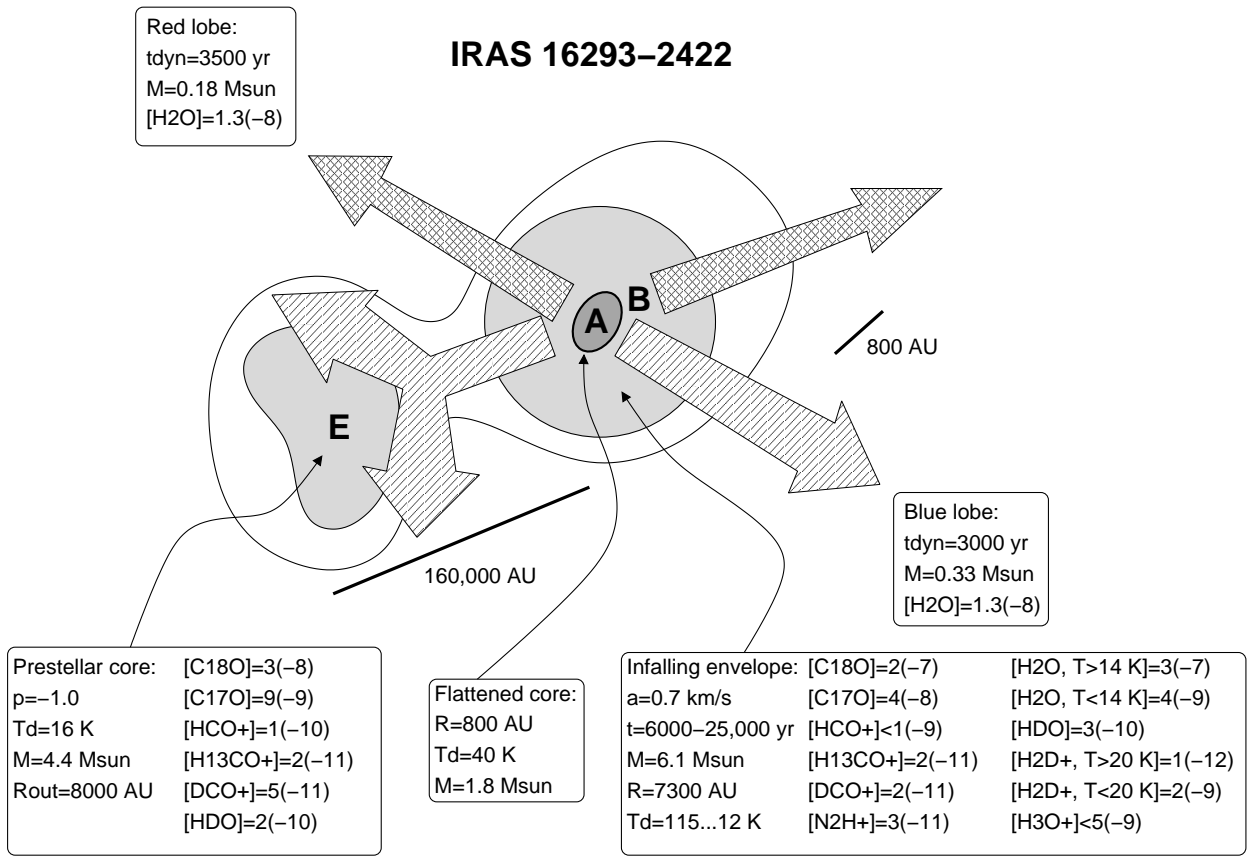}
  \caption{Schematic overview of the IRAS 16293-2422 region showing 
its distinct physical regions with their derived physical parameters, 
and molecular abundances.\label{f17}}
\end{figure}

%%%%%%%%%%%%%%%%%%%%%%%%%%%%%%%%%%%%%%%%%%%%%%%%%%%%%%%%%%%%%%%%%%%%%% FIN

\end{document}